\documentclass[sigconf]{acmart}
\AtBeginDocument{%
 \providecommand\BibTeX{{%
 \normalfont B\kern-0.5em{\scshape i\kern-0.25em b}\kern-0.8em\TeX}}}

\copyrightyear{2024} 
\acmYear{2024} 
\setcopyright{acmlicensed}\acmConference[CHI '24]{Proceedings of the CHI Conference on Human Factors in Computing Systems}{May 11--16, 2024}{Honolulu, HI, USA}
\acmBooktitle{Proceedings of the CHI Conference on Human Factors in Computing Systems (CHI '24), May 11--16, 2024, Honolulu, HI, USA}
\acmDOI{10.1145/3613904.3642219}
\acmISBN{979-8-4007-0330-0/24/05}

\usepackage{comment}

\begin{document}

\title[Eye Gaze Heatmap Analysis of Uncertainty Head-Up Display Designs]{An Eye Gaze Heatmap Analysis of Uncertainty Head-Up Display Designs for Conditional Automated Driving}

\author{Michael A. Gerber}
\orcid{0000-0001-7850-6050}
\affiliation{%
 \institution{CARRS-Q - Queensland University of Technology, msg Systems AG}
 \country{Australia, Germany}}
\email{michael.gerber@msg.group}

\author{Ronald Schroeter}
\orcid{0000-0001-7990-1474}
\affiliation{%
 \institution{CARRS-Q - Queensland University of Technology}
 \country{Australia}}
\email{r.schroeter@qut.edu.au}

\author{Daniel Johnson}
\orcid{0000-0003-1088-3460}
\affiliation{%
 \institution{Queensland University of Technology}
 \country{Australia}}
\email{dm.johnson@qut.edu.au}

\author{Christian P. Janssen}
\orcid{0000-0002-9849-404X}
\affiliation{%
 \institution{Utrecht University}
 \country{Netherlands}}
\email{C.P.Janssen@uu.nl}

\author{Andry Rakotonirainy}
\orcid{0000-0002-2144-4909}
\affiliation{%
 \institution{CARRS-Q - Queensland University of Technology}
 \country{Australia}}
\email{r.andry@qut.edu.au}

\author{Jonny Kuo}
\orcid{0009-0001-6422-4623}
\affiliation{%
 \institution{Seeing Machines Ltd}
 \country{Australia}
}
\email{jonny.kuo@seeingmachines.com}

\author{Mike G. Lenne}
\orcid{0000-0003-1671-6276}
\affiliation{%
 \institution{Seeing Machines Ltd, Canberra}
 \country{Australia}
}
\email{mike.lenne@seeingmachines.com}

\settopmatter{authorsperrow=4}

\renewcommand{\shortauthors}{Gerber, M. A. et al.}

\begin{abstract}
This paper reports results from a high-fidelity driving simulator study (N=215) about a head-up display (HUD) that conveys a conditional automated vehicle’s dynamic “uncertainty” about the current situation while fallback drivers watch entertaining videos. We compared (between-group) three design interventions: display (a bar visualisation of uncertainty close to the video), interruption (interrupting the video during uncertain situations), and combination (a combination of both), against a baseline (video-only). We visualised eye-tracking data to conduct a heatmap analysis of the four groups’ gaze behaviour over time. We found interruptions initiated a phase during which participants interleaved their attention between monitoring and entertainment. This improved monitoring behaviour was more pronounced in combination compared to interruption, suggesting pre-warning interruptions have positive effects. The same addition had negative effects without interruptions (comparing baseline \& display). Intermittent interruptions may have safety benefits over placing additional peripheral displays without compromising usability.
\end{abstract}




\begin{teaserfigure}
 \includegraphics[width=\textwidth]{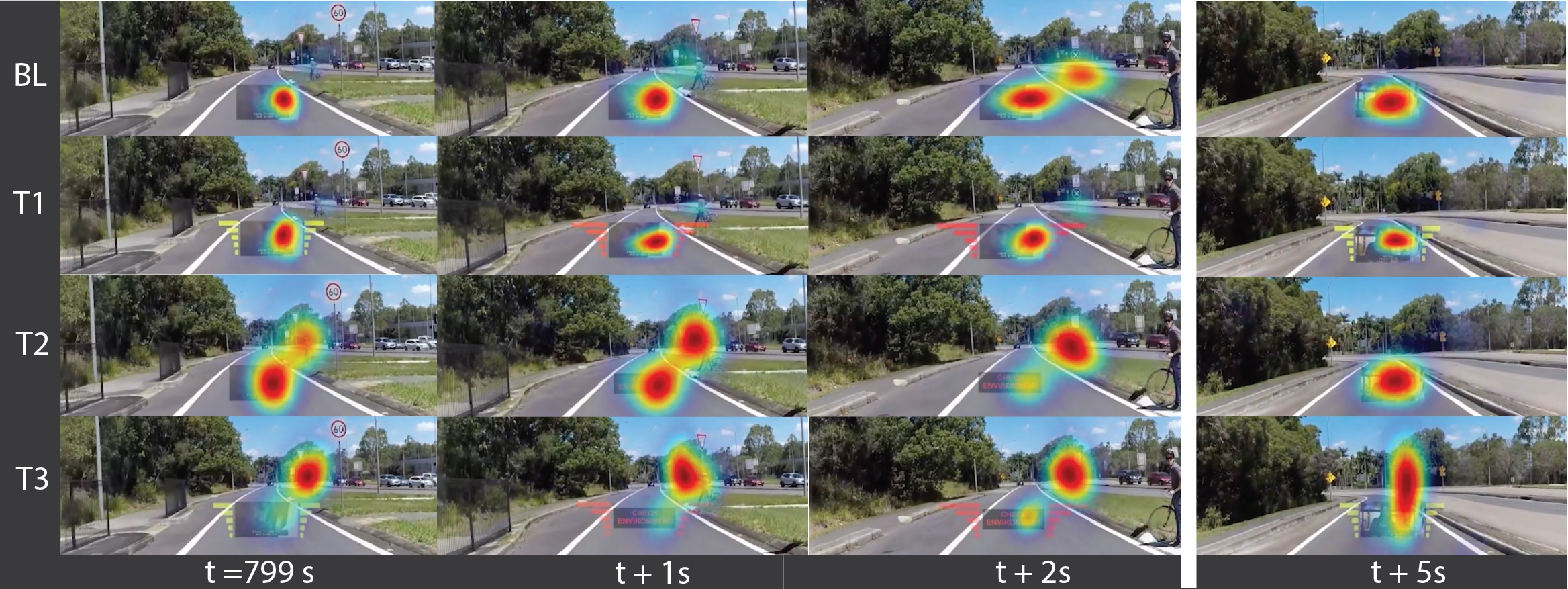}
 \caption{Time-series eye-gaze heat-map of a developing hazard of this between-group automated driving simulator study comparing four head-up treatments. It indicates (details in results) that participants in T2-Interruption and T3-Combination exhibit a more appropriate hazard monitoring behaviour, while Baseline and T1-Display focus on non-driving activities. Additionally, T3 reacts faster than T2 at t=799s, and at t+5s is another monitoring "echo" following earlier interruptions in T3}
 \Description{This figure shows 4 times the same driving scene in a grid. Each of the driving scenes includes a different display intervention that support fallback readiness, described in the intervention section. Overlayed on top of it are the eye-gazes of the participants in form of a heatmap. The details are described in the discussion}
 \label{fig:teaser}
\end{teaserfigure}

\received{14. September 2023}
\received[revised]{12. December 2023}
\received[accepted]{19. January 2024}

\maketitle

\section{Introduction}
The capabilities of automated vehicles are rapidly changing, which in turn changes the nature of the driving task and the roles humans need to perform. The driving task— including monitoring the driving situation, making decisions, and the longitudinal and lateral control of the vehicle—is transitioning from being the sole responsibility of the human driver to being (partially) that of the automated driving system. Due to this transition, human users must monitor the automation and be the backup whenever the automation fails or reaches some limitation. The step-wise transition and shifting responsibilities are defined in a taxonomy from the Society of Automotive Engineers (SAE) \cite{SAE_International2021-ix}.

In \textit{SAE Level 3 (L3)}, also called \textit{conditional automated driving} and the focus of this paper, the driver is—for the first time—freed from supervision duties. SAE L3 driving systems are defined as capable of self-diagnosing poor performance or limitations by themselves and should request the human user to take control when needed, known as the Request to Intervene (RtI) or Take Over Request (TOR). The human user is to \textit{remain fallback-ready at all times} to \textit{safely} take control of the vehicle in a timely manner \cite{Janssen2019-ib, Marberger2018-cm, SAE_International2021-ix}. 
 
This definition poses the following safety conundrum: Without driving-related tasks and responsibilities, human users are enabled and enticed \cite{Naujoks2016-lr, Neubauer2012-qy, Rudin-Brown2004-ez} to do other tasks, often referred to as non-driving related activities (NDRAs). Those NDRAs can degrade performance when taking over control with potential safety implications. Research suggests that even if these NDRAs were not allowed (i.e. made illegal), humans would likely be engaged in them \cite{Naujoks2016-lr, Neubauer2012-qy, Rudin-Brown2004-ez}, which further amplifies this conundrum. Unfortunately, we can see this behaviour in the proliferation of illegal mobile phone use during manual driving \cite{Nguyen-Phuoc2020-zz}.

NDRAs degrade performance when taking over control, especially in safety-critical transitions, due to adverse effects on the user’s \textit{Situation Awareness} \cite{Endsley2017-qd, Liang2021-hk, Wan2018-cu}. Fallback-ready users perform better in this time-limited transition with good situation awareness \cite{Endsley2017-qd, Gold2013-js, Schroeter2016-fj}. As a result, \cite{Gerber2023-lw, Endsley2016-ty, Endsley2019-ac} suggests that the driver should be kept “in the loop” until automation reaches higher levels where no safety-critical transition is required. This implies that the fallback-ready user should maintain situation awareness while acting as a fallback to the automated driving system. To achieve this, we need to support the first level of SA, the perception of a situation, by guiding a user's gaze onto the road despite other tasks. 

This paper evaluates three interventions designed to help the fallback-ready user maintain and balance situation awareness while attending to an NDRA on a head-up display (HUD) during automated driving. Specifically, the design interventions build on research suggesting the benefits of sharing the uncertainty of the automated driving system to the user while they engage in an NDRA \cite{Beller2013-ir, Helldin2013-dq, Kunze2019-gj, Louw2017-tu}

Uncertainty represents the automation’s reliability or confidence in mastering the current driving situation. Consequently, uncertainty also implies the likelihood of the automated driving system potentially failing or reaching a limitation of its designed functions. 

This paper proposes and evaluates various uncertainty-based design concepts to explore the following research question: \textit{How do different uncertainty design intervention concepts a) influence the eye-gaze behaviour of fallback-ready users while engaging in an everyday non-driving related activity, and b) impact usability?} It further discusses the implications regarding the future design of such interventions to balance safety and usability. 

This work contributes significantly to understanding users in the context of SAE L3 automated driving. We analyse user’s gaze behaviour using heatmap analysis and self-reported system usability using the System Usability Scale in relation to using different HUD design artefacts. We discuss broader design implications to keep future users of SAE L3 automated vehicles safer through improved head-up display designs and for balancing safety and user experience in safety-critical environments—where attention matters.

\section{Related Work}
\subsection{Attention Management Systems}
The safety conundrum of vehicle automation mentioned above calls for systems designed to manage better the user’s attention between the driving environment and an NDRA. Such systems can be referred to as attention management systems to optimise situation awareness, trust, fallback-readiness and/or take-over performance \cite{Endsley2016-ty, Park2019-ad} and have also been developed outside the driving domain where multiple tasks compete for attention \cite{Anderson2018-bt, Janssen2019-af}. 

Attention management systems come in various forms and may include alerts based on driver state, such as extended glances off-road \cite{Llaneras2017-sw}, or gamification \cite{Schroeter2016-fj, Zimmermann2018-yd, Steinberger2017-qp} to draw a user’s attention to the driving scene through an engaging driving related activity. 

In this paper, we explore an attention management concept that sits alongside an engaging non-driving related activity (NDRA) — creating a dual-task competition — and aims to shift the user attention in an active (cf. alerts) and targeted way (i.e. interrupt the NDRA when a situation is more likely to require it). This targeting may focus on factors that influence the performance (time and quality) of a driving manoeuvre, which includes a) the state and driver capabilities of the human driver, b) the vehicle, including its automation, interfaces and controls, and c) the driving environment or traffic situation \cite{Marberger2018-cm, Radlmayr2014-ar}. Our work focused on the latter, the driving environment's complexity, or better, the potential uncertainty of an automated driving system relative to the traffic situation, as discussed in turn. 

\subsection{Uncertainty in Automated Driving}
Uncertainty represents the automation’s reliability or confidence in mastering the current driving situation. This uncertainty may be presented to the user—with the aim to maintain fallback-readiness — in different modalities. Several studies have investigated different design methods to visualise and use the function or information of uncertainty to maintain fallback readiness. Finger \& Bisantz \cite{Finger2002-pu} report, based on the results of two studies, that uncertainty information visualised as graphics is as effective as verbal or numerical communication for the decision-making task. At the same time, Walch et al. suggest multi-modal notifications of handover scenarios lead to better performance \cite{Walch2015-na}. Our paper focuses on generating knowledge that aids the optimisation of the visual modality, which may be complemented by other modalities in the future.

Several independent works demonstrated the utility of visualising uncertainty to improve various safety measures in SAE L3 automated driving. Kunze et al. \cite{Kunze2019-gj} evaluate an uncertainty visualisation in the form of a numeric heartbeat visualisation: The higher the heartbeat, the higher the uncertainty of the automated system. Based on the driving simulator experiment that manipulated uncertainty through an emerging fog, the authors suggest that uncertainty communication leads to more appropriate attention allocation strategies for fallback-ready users. This may lead to higher situation awareness and better takeover performance. 

Reports on broader benefits of visualising uncertainty, such as trust in and acceptance of automation, is still debated. Beller et al. \cite{Beller2013-ir} use anthropomorphic smileys performing hand gestures to visualise uncertainty. This increased trust in and acceptance of the automation in their driving simulator study. Helldin et al. \cite{Helldin2013-dq} have observed the opposite in lower trust ratings. That said, they argue that their display—an uncertainty visualisation with a graphical representation of the automation’s ability in 7 bars—actually leads to an overall better trust \textit{calibration} even if trust is lowered.

There is little research on incorporating the NDRA in the uncertainty design concept. Neither Helldin et al. nor Beller et al. incorporated a \textit{natural} NDRA in their uncertainty designs or study setup. Kunze et al. \cite{Kunze2019-gj} report that their heartbeat design impeded participants from engaging in the NDRA., which may cause the user to turn off the visualisation, negating its benefit. This research investigates a less intrusive method to convey the vehicle's automation uncertainty by incorporating the NDRA in the uncertainty design and vice versa.

Regarding visualisation guidelines, Kunze et al. \cite{Kunze2018-do} identify the best suited variables to visualise uncertainty in the context of automated driving and augmented reality displays. They explored colour-based variables, such as hue, animation-based variables and other dimensions (e.g.position, size, value, orientation). Their recommendation to use colour changing, hue and animation informed some of our intervention design (see Intervention Design). In addition, we incorporated Helldin et al.’s design of increasing bars that suggested better trust calibration \cite{Helldin2013-dq}.

\subsection{NDRA Interruptions During Automated Driving}
SAE L3 marks a critical milestone concerning NDRAs in that the automation is expected to be so advanced that an NDRA is no longer considered secondary to the primary driving task but rather a competing parallel task \cite{Janssen2019-ib}. This creates a dual-task condition for which we need to understand better how people interrupt one task (NDRA) to attend to the other (monitoring driving automation) so we can start to better understand potential safety implications \cite{Janssen2019-ib, Marberger2018-cm}. 

Past studies often used standard tests for assessing the engagement in NDRAs, such as n-back tasks \cite{Naujoks2018-ve}. Those tests typically provide consistency in controlled experimental settings to quantify measurable properties of specific demands. The contrived nature of such tasks means that motivational aspects, particularly intrinsic ones, are considered less, even though they are likely to play a critical role in automated driving and NDRAs of more intrinsic interest to the user. To that end, motivational aspects have been shown to impact the take-over performance \cite{Janssen2019-ib, Marberger2018-cm, Naujoks2018-ve}. 

The NDRA of watching videos has been identified as a likely intrinsically motivated activity during automated driving \cite{Klobas2018-oi}. Gerber et al. \cite{Gerber2020-oh} investigated users’ NDRA interruptions in this scenario. They compared mobile phones vs HUDs as the output modality and found HUDs to have potential safety benefits. However, they did not attempt to improve the HUD design to optimise gaze and interruption behaviour. This paper aims to study (and design for) NDRA interruptions, comparing gaze behaviour in different HUD design concepts. 

The nuances of gaze behaviour in this context have been studied by Louw and Merat \cite{Louw2017-tu}. They demonstrated that the vertical gaze was more dispersed with high uncertainty. In their work, uncertainty was not communicated via Human-Machine Interface (HMI) but externally manipulated by increasing fog levels in the driving simulation. However, their results indicate that participants adjust and optimise their gaze strategies based on perceived increasing uncertainty. As the study did not include an NDRA—or dual-task competition—it is unclear how the motivational conditions of a desired or entertaining task could negatively impact the perception of increased uncertainty and, therefore, the adjustment of gaze strategies. 

Regarding managing situations where two tasks are in competition and parallel processing is impossible, Wickets et al. \cite{Wickens2004-ga} suggest that switching these tasks might be necessary. Following this, we propose methods to interrupt the NDRA based on a function of uncertainty to promote task interleaving with the monitoring task of automated driving. McFarlane \cite{McFarlane2002-eu} suggests the following methods to that end: a) “immediate” interruption with unpredictable behaviour and without pre-warning; b) “negotiated” interruption, where the user is suggested to interrupt but in control of the task switch; c) “mediated” interruption, where the system interrupts one task independent from its state but announces the task switch; and d) the “scheduled” interruption in a regular pattern. Since our approach incorporates uncertainty, “scheduled” is not appropriate, and we focus on the first three methods, as discussed in detail in the following section.

\section{Uncertainty Design Interventions}
This study assesses a novel HMI conveying a vehicle's current “uncertainty” of the driving situation. Uncertainty is modelled on different levels (e.g. low, medium, high), which correspond to the likelihood of an impending TOR, i.e. if uncertainty is high, the likelihood of a TOR occurring is also high, provided the situation escalates further. Conceptually, this follows the principle of graceful degradation, aimed at preparing users for a potential task switch \cite{Hancock2019-qf, Janssen2019-ib}.

We created two uncertainty concepts, one \textit{visual graphics/ animation}-based and the other \textit{interruption}-based, resulting in three interventions, one for each concept, plus a combination of both. These are inspired by McFarlane's interruption treatments \cite{McFarlane2002-eu}. Note that in addition to the descriptions of those three interventions below, we have also published a video demonstration\footnote{https://www.youtube.com/watch?v=ZuFTiiR8wR0, retrieved January 2024} \cite{Gerber2021-om}:

\subsection{Uncertainty Intervention 1: Display}
The first intervention—\textit{Display}—intends to visually display uncertainty information to the fallback-ready user via animated graphics (as suggested by Finger \& Bisatz \cite{Finger2002-pu}). The visualisation is placed right next to the NDRA content screen for watching a video over a HUD (see HUD section under Methodology). Visualising uncertainty to draw attention back to monitoring reflects the “negotiated interruption” of McFarlane's study design \cite{McFarlane2002-eu}. Previous research suggests that NDRAs encouraging drivers' eye-gazes towards the road centre may counteract “being out-of-the-loop” or improve situation awareness \cite{Gerber2020-oh, Louw2017-tu}. 

Our uncertainty intervention display applies four of Kunze et al.'s \cite{Kunze2018-do} design principles, which they identified as having the strongest subjective perceived impact on uncertainty: the hue—in our case, visualised from green over yellow to red—and the size, movement and direction of the animated bars. All of these resulted in a wing-shaped design (Figure \ref{fig:interventions}, T1), which we named Guardian Angel.

\subsection{Uncertainty Intervention 2: Interruption}
The second uncertainty intervention — \textit{Interruption} — is simply the interruption of the visual aspect of the NDRA and follows the design of the “immediate interruption” \cite{McFarlane2002-eu}. We drew inspiration from the classical task switch research \cite{Brumby2019-hr} to make room for regular soft task switches of monitoring instead of having a hard task switch to the safety-critical driving task resumption. Switching between tasks takes time to recover and might lead to errors \cite{Salvucci2009-gq}. Wickens et al. state that if parallel processing is not possible, switching between tasks can be good or even necessary \cite{Wickens2004-ga}. 

Therefore, the idea behind Interruption is to take the visual stimuli of the NDRA temporarily away so that it brings the visual stimuli of the driving environment to the fore. This may be beneficial for gaining situation awareness until the vehicle’s uncertainty is overcome and the NDRA returns or until the TOR is issued if uncertainty continues to grow beyond vehicle capability. 

The way we implemented this in practice for this study was to replace the video content in the HUD with a static “check environment” message (Figure \ref{fig:interventions}, T2) when the uncertainty was high. The audio of the video continued over the speakers. When uncertainty decreased again, the visual video content would automatically reappear.

\subsection{Uncertainty Intervention 3: Combination (Display \& Interruption)}
\begin{figure}[ht]
 \centering
 \includegraphics[width=\linewidth]{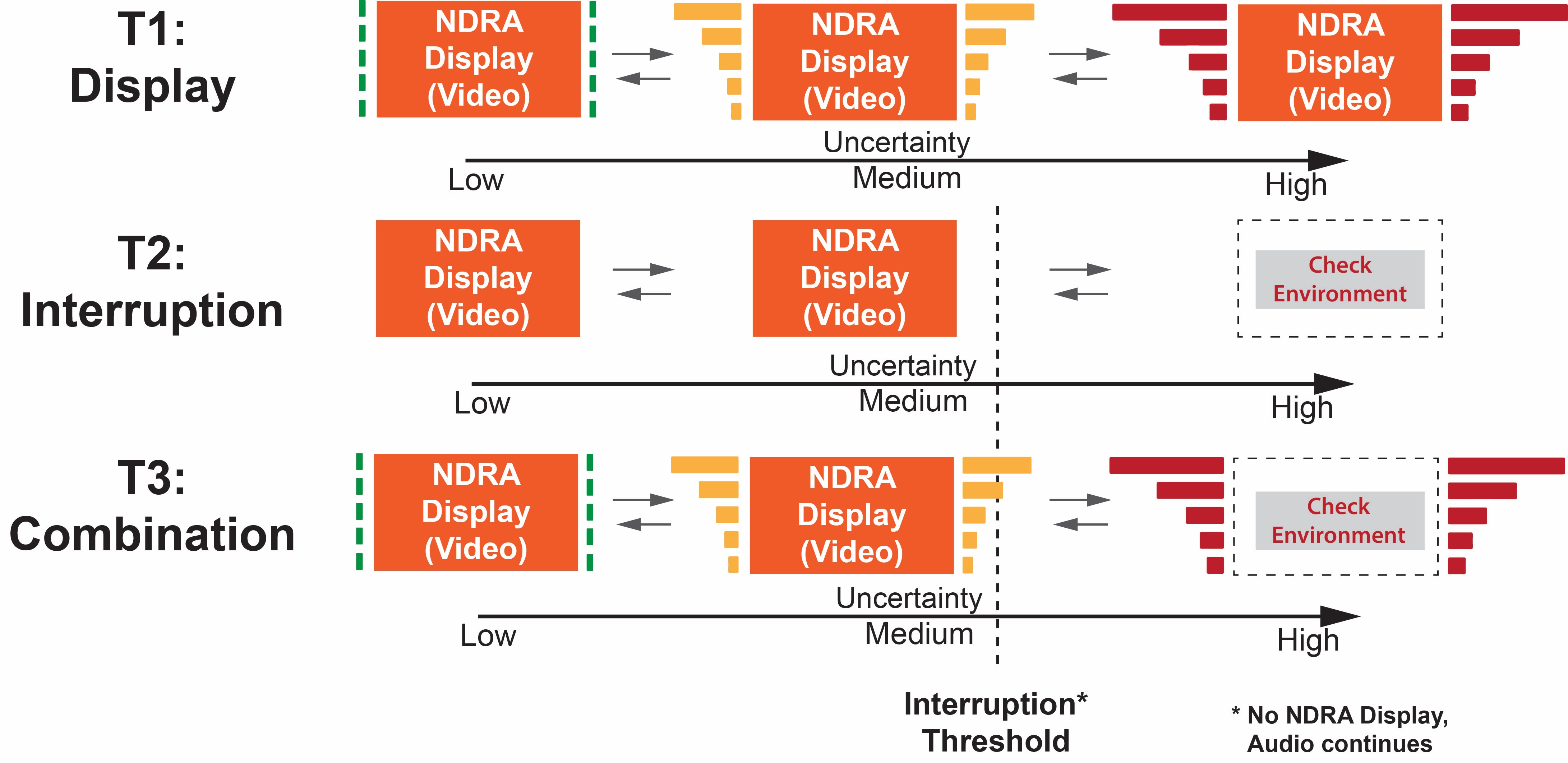}
 \caption{Three Uncertainty Interventions are tested against a baseline (just a display) in this study design. Intervention 1 (top) adds to the display the “Guardian Angel”, a visualisation of uncertainty with colour and size-changing bars. Intervention 2 (centre) interrupts the NDRA by increasing uncertainty. Intervention 3 combines Intervention 1, the visualisation and Intervention 2 the Interruption. }
 \Description{3 Interventions Designs are visualised, the first a display with bars on the side that indicate the certainty of the vehicle from green to red, and the second intervention is just the display which disappears at high uncertainty. The third intervention is a combination of both}
 \label{fig:interventions}
\end{figure}

The third intervention—\textit{Combination}—combines aspects of the first two and follows the principle of “mediated interruption” \cite{McFarlane2002-eu}. It provides the visual, wing-shaped design feedback of the intervention \textit{Display} and replaces the video content with the static “check environment” message corresponding with \textit{Interruption} (Figure \ref{fig:interventions}, T3). This concept follows a principle described by Endsley \cite{Endsley2015-dt} in the context of data-driven vs. goal-driven data processing. While the data-driven processing (i.e. automation uncertainty visualisation) is a bottom-up approach of conveying the driving situation to the fallback, the actual monitoring represents the goal-driven approach (top-down) where the user assesses the situation by themselves. Combining both approaches would force the user to use the “\textit{natural}” human alternating information processing method by switching between both models \cite{Endsley2015-dt}.

\section{Methodology}
In the following, we describe the methodology to assess the behavioural changes of the described interventions. Automated driving, a head-up display and the interventions were simulated using a video simulation in an Advanced Driving Simulator equipped with a driver monitoring system to measure gaze behaviour. Post-drive subjective self-report measures were supported by freeze-frame callbacks transporting participants back to particular driving situations they just experienced. In a between-group study with N=215 participants, we then investigated gaze behaviour and usability. The details of the methodology are discussed in turn. 

\subsection{Experiment Setup}
\subsubsection{Apparatus}
The study took place in the Advanced Driving Simulator of our research centre. The simulator (custom-built by Oktal SYDAC) includes a right-handed Holden Commodore vehicle on a 6-degree-of-freedom (6DOF) motion platform. The motion platform elevates the vehicle into a consistent and calibrated operating position, which remained static in this position during the simulation throughout the study. 

The front of the simulator has three 3x4m video screens wrapping around 180 degrees to simulate the forward view. The rear view is simulated through three small LCDs in place of the rear mirrors. In this study, the simulator also included a digital speedometer in the instrument cluster behind the steering wheel in the form of a 6.4-inch OLED display, a navigation display above the centre console in the form of a 5.5-inch LED display, a remote-controlled self-moving steering wheel from SensoDrive, and a surround audio system with a subwoofer under the driver's seat for vehicle vibrations and driving sounds.

Seeing Machines’ Driver Monitoring System (DMS) was used to record head and gaze tracking (at 30 frames per second). Its camera was firmly secured above the instrument cluster to ensure it did not move between participants and had an unobstructed view of the participant’s head. 

The DMS was synced temporally and spatially. Temporally, the DMS’ CAN-Bus outputs (incl. its frame numbers) were recorded with a sensor fusion software RTMaps from Intempora \footnote{\url{https://intempora.com/products/rtmaps/}, retrieved September 2023}, alongside the network messages of the automated driving simulation (see below). This resulted in an overall synchronisation accuracy of \textpm two frames between participants. Spatially, the DMS software \footnote{\url{https://seeingmachines.com/products/automotive/}, retrieved September 2023} allows for creating a 3D scene defining all areas of interest within the Advanced Driving Simulator (e.g. front screen, read mirror, navigation screen) and their exact relative position to the DMS camera in the operating position. A calibration for each participant was not required since the DMS self-calibrated each participant. However, a control sequence to look at specific dots was used in the introduction procedure for backup and validation. 

\subsubsection{Automated Driving Simulation}
An immersive video simulation method was used to simulate conditional automated driving capabilities while providing the visual richness of the natural world \cite{Gerber2020-oh, Li2020-yb, Li2021-ff, Haeuslschmid2017-hx}. To create the video simulation, the research team drove a (non-automated) car on local roads near the lab where the study took place. The car was equipped with multiple cameras and data loggers to record a) six driving videos (three for the front screen and three for the mirrors), b) a GPS/navigation video for the navigation display, c) speed (to show current speed on the digital speedometer, d) steering wheel angles (to “replay” steering wheel movements), and e) the interior audio during the drive to play back driving sounds via the audio system. Overall, this provides a high-fidelity automated driving simulation in an environment that local participants are highly familiar with, albeit with some limitations (see Limitations section).

The simulated drive (visualised in Figure \ref{fig:NDRAEngagementGraph}) represents a typical trip in the city where the study took place during non-rush hours in diverse traffic environments that drivers would experience during a typical daily commute. Specifically, the trip started in the inner city (characterised by low speed, dense traffic) to the suburbs (less dense traffic, less turns, low speed) via a major arterial highway road (multi-lane highway driving, higher speed, medium traffic). During this sequence, there were various typical driving events (cf. with diamonds in Figure \ref{fig:NDRAEngagementGraph}) such as (E1) a right turn at the beginning of the drive, (E2) standing at a traffic light, (E3) driving through an underpass with blinding light, (E4) a long left turn at the city centre with dense traffic, (E5) merging on the highway, (E6) exiting the highway and (E7) an emerging hazard in the form of a pedestrian threatening to cross the suburban road without looking at the approaching vehicle. 

\subsubsection{Quantifying Uncertainty}
In contrast to more controlled driving experiments, the naturalistic nature of our video clip did not allow for a straightforward quantification of the dynamic \textit{uncertainty} of the driving situation. We, therefore, used a computerised tool and a separate small study to create the uncertainty metric needed for the study presented in this paper.

\begin{figure}[ht]
 \centering
 \includegraphics[width=\linewidth]{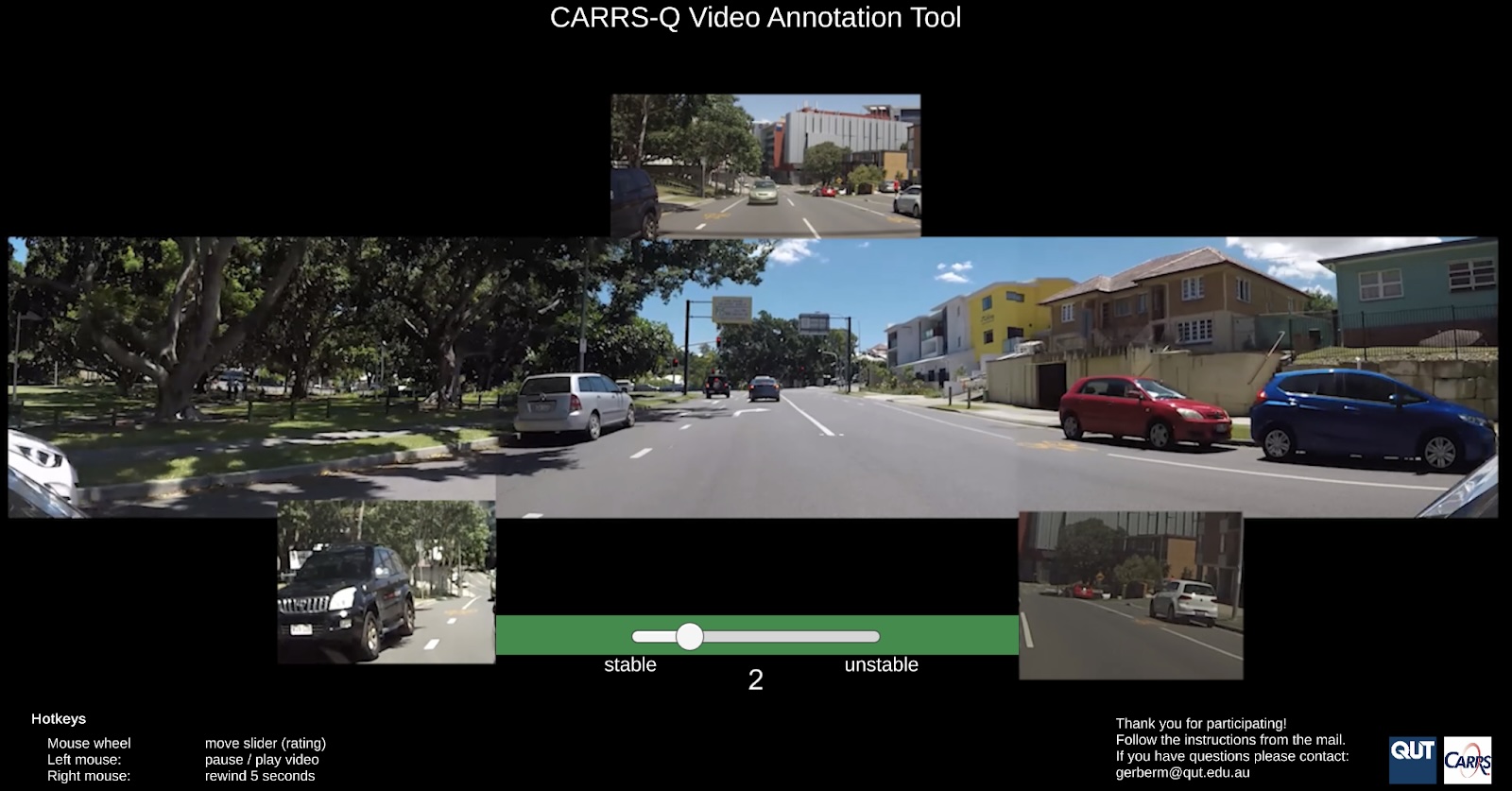}
 \caption{Remote data collection tool for quantifying the driving environment by experts. They rated from stable (0) to unstable (9) by rotating their mouse wheel. The colour of the bar changed from green over yellow to red as additional feedback}
 \Description{The Image shows a driving environment recorded from a vehicle, including the views of the mirrors. At the bottom is a scroll bar that lets the participants quantify the driving situation}
 \label{fig:VideoAnnotationTool}
\end{figure}

We developed a tool in Unity \footnote{\url{https://unity.com}, retrieved September 2023} (Figure \ref{fig:VideoAnnotationTool}) to replay the 180-degree and mirror videos, and a mouse wheel function to continuously rate (assign a numerical value to) the entire 14-minute drive. We then recruited N=18 international automated driving domain experts from Computer Science, Engineering, Psychology, Human Factors, and Road Safety. Using the tool and while watching the video in real time, they were asked to assign a numerical value of their perceived stability/complexity from one perspective they could self-select: a) the stability of the driving environment (N=5), b) machine sensory perception (N=4), c) machine decision making (N=4), or d) human decision making (N=5). The experts received 20 AUD in vouchers in appreciation of their time, which the university’s ethics board approved with the approval number 1800000205.

The experts’ response values, collected at a sampling rate of 10Hz, were normalised. The mean value was cut off at the bottom when the vehicle was in a less complex situation (i.e. more stable and less uncertain, e.g. standing at a traffic light such as event E2 in Figure \ref{fig:NDRAEngagementGraph}) and at the top when the situation was rated most complex (i.e. more unstable and more uncertain, e.g. merging several lanes, the emerging hazard such as events E5 and E7 in Figure \ref{fig:NDRAEngagementGraph}). The resulting uncertainty value throughout the drive is visualised at the top of Figure \ref{fig:NDRAEngagementGraph}. 

 Overall, the result set was deemed an appropriate basis for approximating the uncertainty concerning the driving situations and simulating the experience of a potential uncertainty-based HMI, including erratic and rapid escalations and de-escalations. For the two interruption interventions, the research team picked one constant threshold (0,555) applied over the entire graph that would then trigger multiple interruptions of varied lengths throughout the drive (cf. Figure \ref{fig:NDRAEngagementGraph}).

\subsubsection{Simulating HUD for NDRA \& Uncertainty Intervention}

The HUD—for displaying an NDRA (watching video) and the uncertainty display—was simulated by editing the front screen video and overlaying the HUD elements.

\begin{figure*}[ht]
  \centering
  \includegraphics[width=\textwidth]{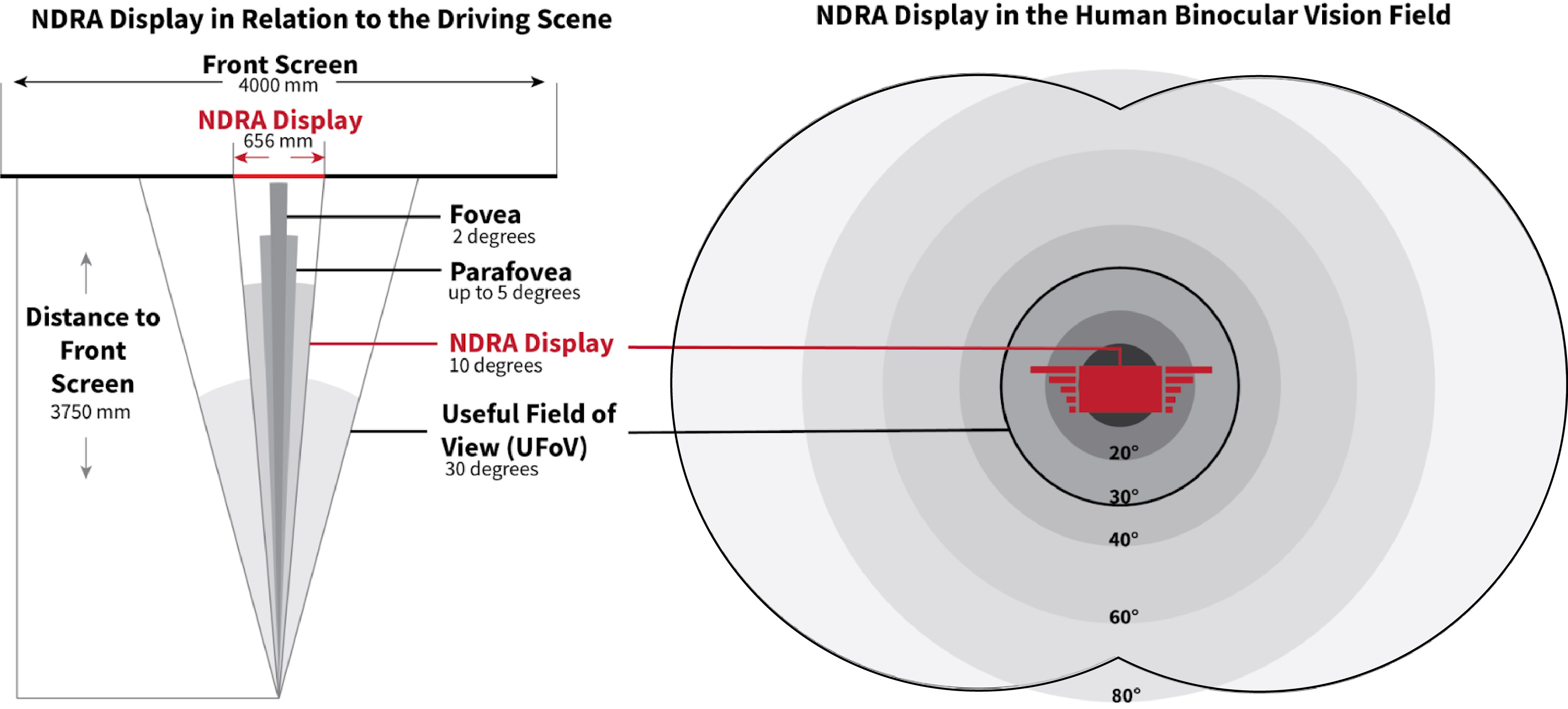}
 \caption{Size of NDRA Displays is designed to enable the peripheral vision of driving scene and uncertainty display based on \cite{Wolfe2017-dq}}
 \Description{Visualised are the different field of views a human can see best. This radia are then projected on a wall, where the size of the display is visualised}
 \label{fig:FoveaDisplayPositioning}
\end{figure*}

The size of the NDRA’s video content display is based on the distance from the drivers' eyes to the front screen (3750mm) and the Useful Field of View (UFov) of 30 degrees (Figure \ref{fig:FoveaDisplayPositioning}). Since in transportation research, 30° and peripheral research 20° \cite{Kotseruba2021-md} are suggested, the content display is set to the size of 10° to provide space for the peripheral display area from 5° to 10° from the Fovea centre. This display size was tested in an Advanced Driving Simulator when piloting the study with over 10 colleagues external to the project to ensure a reasonable size to watch video content. During pilot studies that involved colleagues, we adjusted the size of the screen to 656mm wide, which subjectively felt to provide both a sufficient overview of the driving context and allow them to read subtitles that were shown in the video (in other words, balancing peripheral and focal vision). 

\begin{figure}[ht]
 \centering
 \includegraphics[width=\linewidth]{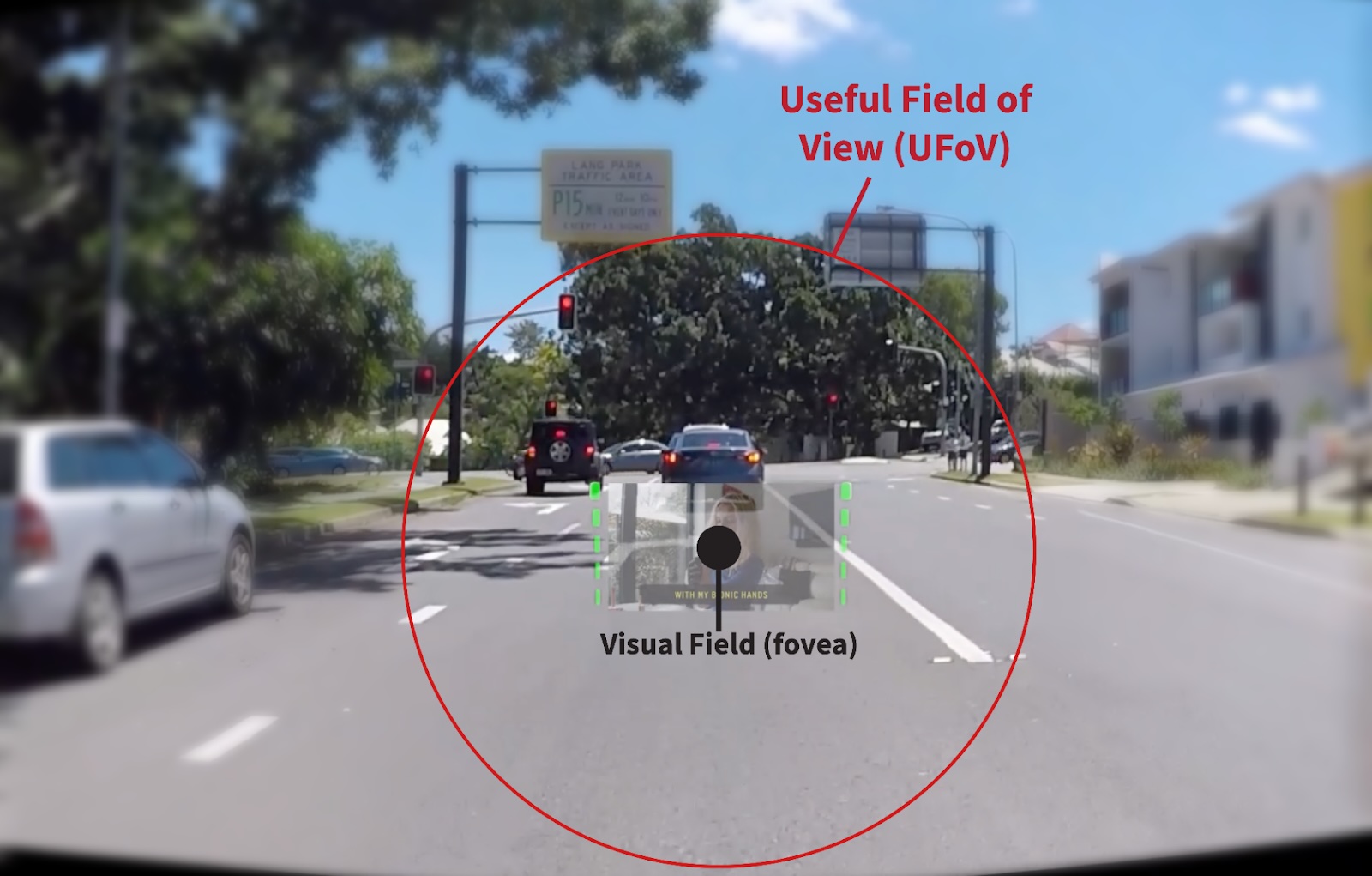}
 \caption{The NDRA display in the simulator setup driving scene with the useful field of with the loss of resolution in the periphery according to Geisler \& Perry model \cite{Geisler1998-jc} (note: blur added for illustration purposes only)}
 \Description{The image shows the driving scene. A radius in the centre visualised the useful field of view. Everything outside of it is blurred. In the centre of the UFOV is the display intervention placed, including a bigger black dot that represents the visual field or fovea of a human}
 \label{fig:UFOVVisualisationOfSimulation}
\end{figure}

The visualisation of our uncertainty display “Guardian Angel” effectively ended up within 20° from the Fovea centre and in the peripheral of the NDRA (Figure \ref{fig:FoveaDisplayPositioning}). Apart from the size, we set the entire HUD to an opacity of 70\% to enable gazes through the display, for example, to perceive line markings (Figure \ref{fig:UFOVVisualisationOfSimulation}). This value was used in a previous simulator study \cite{Gerber2020-oh, Li2020-yb, Li2021-ff}, in which participants perceived it as a reasonable trade-off of being able to engage in the NDRA and monitoring the driving environment. 

\subsubsection{Simulating Dual-Task Competition}
In SAE L3 automated driving \cite{SAE_International2021-ix}, an NDRA competes for the user’s attention against driving-related tasks, such as monitoring the driving environment and/or remaining ready to continue driving after a Take-Over Request (TOR). Given our experiment setup—in particular, the lack of risk to participants when not monitoring—we needed to incentivise attending to the driving task, in this case, checking the environment while also paying attention to the video content of the NDRA.

\begin{figure}[ht]
 \centering
 \includegraphics[width=\linewidth]{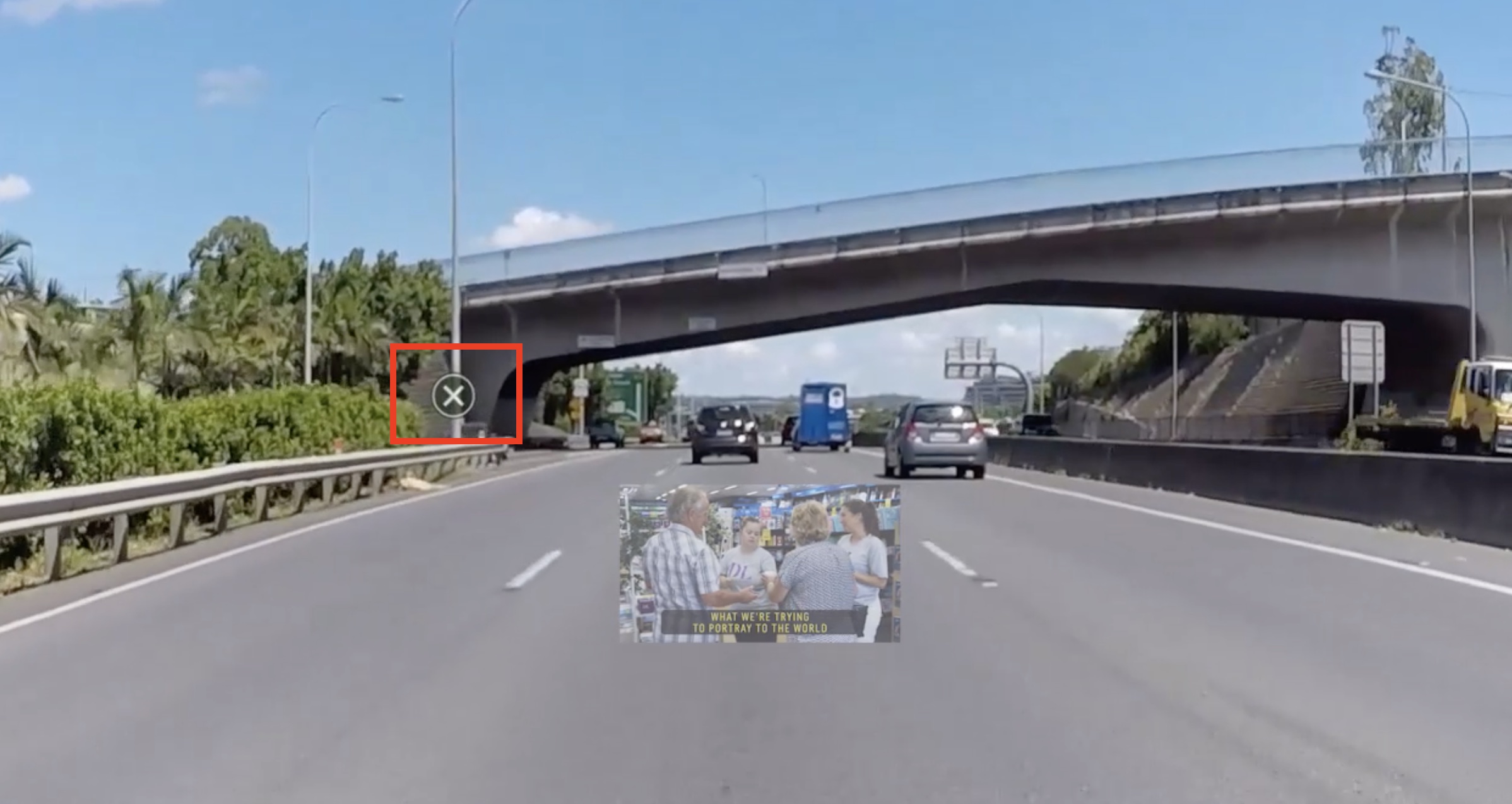}
 \caption{A representation of a superimposed marker in the driving environment that the participants were asked to click respond to (red box for paper presentation purpose only)}
 \Description{The image shows the driving scene. A symbol that looks like a street sign is marked with a red box to find the virtual marker the participants need to react to better}
 \label{fig:MarkerInDrivingScene}
\end{figure}

To simulate this dual-task competition, we edited the driving videos to superimpose round street signs (markers) with a white X on a green background in the driving videos in such a way that they would blend in realistically within the environment (using CGI as if they were real signs, Figure \ref{fig:MarkerInDrivingScene}). 

To further amplify the dual-task competition, we instructed participants a) to monitor the environment and click a button in their hands every time they noticed these signs, b) to pay close attention to the video clips of the NDRA—a pre-selected playlist of 13 widely unknown one-minute documentaries from “60-second docs”\footnote{\url{https://www.youtube.com/c/60SecondDocs}, retrieved September 2023}—to complete an NDRA Recall Questionnaire after the drive about the content following the drive (see below), and c) to try to score well in noticing the markers and answering video content questions, but to prioritise the first. In the analysis, the marker clicks and questionnaire data afforded the checking for significant differences between groups, but this was not the main purpose.

\subsection{Study Design}
The study used a between-subject design to mitigate the effects of fatigue and learning effects (since all videos were the same) with four groups : A baseline (BL) without uncertainty intervention, but with the HUD playing back the video for the NDRA while driving; and three treatment groups analogous to the three uncertainty interventions discussed earlier: T1-Display featuring the “Guardian Angel” visualisation, T2-Interruption, T3-Combination (Figure \ref{fig:interventions}). Our experiment setup resulted in each participant/group experiencing the same drive and watching the same NDRA video content on the HUD at the same time relative to the drive. However, the groups differed in the uncertainty treatments (see Uncertainty Design Interventions). 

To explore the effects of the uncertainty design interventions on the visual task-switching behaviour within the simulated dual-task competition, we used Seeing Machines' Driver Monitoring System (DMS) to track head/eye-gaze. The sampling rate was set to 30 FPS to align with the frame rate of the driving video. From this data, we were particularly interested in observing differences in gaze patterns between the groups as perception relates to the first level of situation awareness \cite{Endsley2016-ty}.

To better understand such systems' potential influence on end-user acceptance, we included a measure of usability in the form of the System Usability Scale (SUS) \cite{Finstad2010-pp, Klug2017-xk} using all items. Instead of a 5-point Likert scale, a 7-point Likert scale was used because of the higher likelihood to reflect the respondents' true subjective evaluation \cite{Finstad2010-pp, Klug2017-xk} and then interpolated to the 100-point SUS scale.

Lastly, we administered the NDRA Recall Questionnaire — a multiple-choice questionnaire — asking 13 questions about the content of the 13 documentaries and providing one correct answer, two wrong answers, and the option to select “I don’t know, I need to guess”.

\begin{figure*}[ht]
  \centering
  \includegraphics[width=\textwidth]{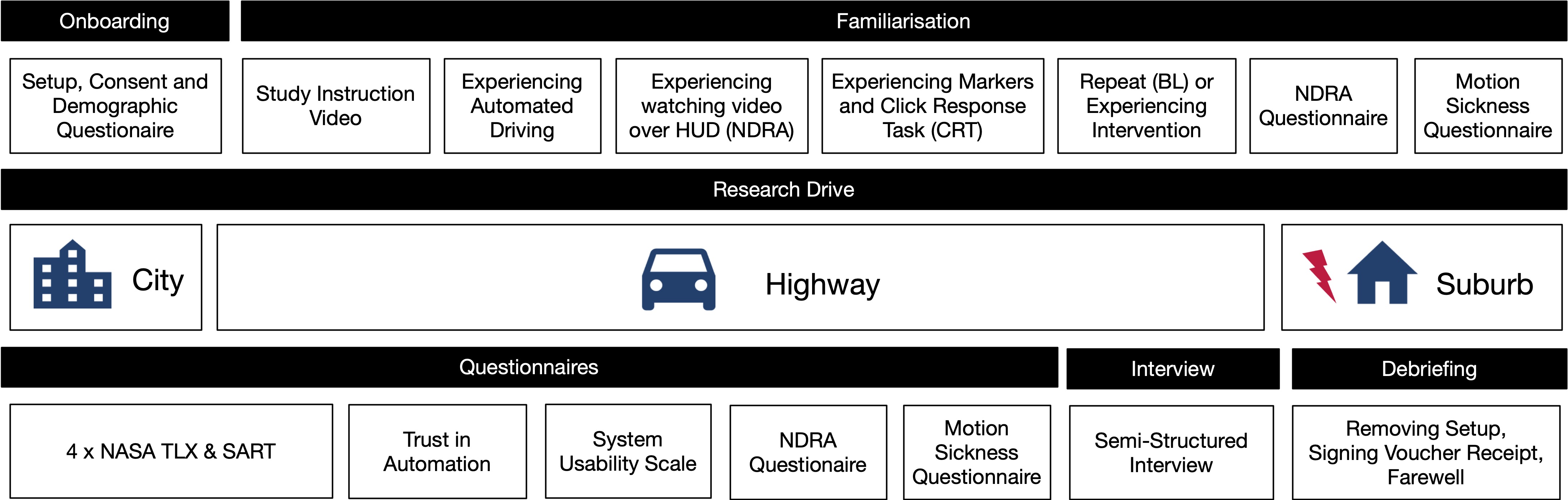}
 \caption{All phases of the visit of a participant starting with onboarding, the familiarisation, the actual data collection research drive, various questionnaires, a semi-structured interview and the debriefing at the end}
 \Description{This Figure visualised the phases of the study, which consist of onboarding, familiarization, research drive, post drive questionnaires, interview and debriefing}
 \label{fig:StudyProceedurePhases}
\end{figure*}

\subsection{Participants}
N=215 (106F, 108 M, 1 other) participants attended this experiment, aged between 20 and 76 years, with an average age of 41.98 and an SD of 13.47 years of age (see Table 1). All participants held a valid and so-called ‘open’ driver's licence \footnote{ An open driver’s licence means that under local jurisdiction a certain minimum driving experience relative to age can be assumed}. They were primarily recruited over targeted social media advertising in the local communities, ensuring familiarity with the local main arterial roads used in the simulation. Additionally, participants were required to speak English as their mother tongue, as this was the language of the experiment instructions and video clips of the NDRA. They were asked to sleep well the night before and bring hearing and visual aids if required. 

The N=215 participants were randomly distributed and balanced over the four study conditions as outlined in Table \ref{tab:Participants} below to mitigate individual differences affecting the results between groups:

\begin{table}[ht]
\resizebox{\linewidth}{!}{
{\scriptsize
\begin{tabular}{lcccc}
 & \textbf{Baseline} & \textbf{Treat 1} & \textbf{Treat 2} & \textbf{Ttreat 3} \\
\textbf{Participants} & N=55 & N=52 & N=54 & N=54 \\
\textbf{Female} & N=27 & N=26 & N=27 & N=27 \\
\textbf{Male} & N=27 & N=26 & N=27 & N=27 \\
\textbf{Other} & N=1 & N=0 & N=0 & N=0 \\
 & \multicolumn{1}{l}{} & \multicolumn{1}{l}{} & \multicolumn{1}{l}{} & \multicolumn{1}{l}{} \\
\textbf{Mean Age} & 39.66 & 40.16 & 44.29 & 43.07 \\
\textbf{Std. Dev.} & 12.8 & 12.61 & 14.36 & 13.35
\end{tabular}%
}}
\caption{Distribution of participant demographics across experimental conditions}
\label{tab:Participants}
\end{table}

The study was conducted following the national Code for the Responsible Conduct of Research and approved by the university’s ethics board with the approval number 1700000425. In appreciation of their time, participants received vouchers as compensation of \$20 for finishing the familiarisation and another \$50 for finishing the data collection drive. 

\subsection{Experiment Procedure}
Figure \ref{fig:StudyProceedurePhases} provides an overview of the experiment procedure, the different phases and building blocks:

\textbf{Onboarding:} Upon arrival at the lab, participants were given general safety instructions for using the Advanced Driving Simulator and allowed to ask any questions; they were then offered to sign consent to participate in the study. Participants were offered something to drink while being equipped with physiological sensors (Biopac BioNomadix EDA and ECG to assess arousal levels during the drive, not reported in this gaze-focused paper). During this time, they filled out the demographic questionnaire on a tablet and refreshed their memory with a printed slideshow about the driving simulator, the safety instructions, the automated driving, and their tasks in this study. Participants were then guided to the vehicle, where they received last explanations of using vehicle controls, such as adjusting the seat position or the fresh airflow. 

\textbf{Familiarisation:} The familiarisation started with a narrated 5:30 min \textbf{introduction video} on the simulator projection screens. It explained the simulation facility, including final safety instructions, such as fire safety procedures and emergency power off. During these safety instructions, the vehicle on the motion platform was lifted to the fixed operational height. It was followed by a study introduction, which explained the automated driving functions for a homogeneous understanding of SAE L3 and their duties to remain fallback-ready. The introduction finished with demonstrating how a TOR would be requested via the vehicle HMIs. This occurred while the vehicle was waiting at a traffic light. 

A detailed 9-minute \textbf{familiarisation drive} followed seamlessly when the traffic light turned green. The automated driving mode was engaged, and participants experienced the first 4 minutes of automated driving without any other tasks and to "settle in". After 4 minutes, the narrator's voice from the introduction video first introduced the NDRA, with the HUD starting to show the video clips. Second, the NDRA display was removed to introduce the markers and click task. Third, the NDRA display was returned, and participants were instructed to do both, watch the video clips and click for markers. Lastly, depending on their respective treatment, participants either repeated the previous practice round in the baseline group or received an explanation according to their intervention group, followed by a practice round. 

After completing the familiarisation drive, a \textbf{practice NDRA Recall Questionnaire} was provided (see Simulating Dual-Task Competition) as well as a Motion Sickness Questionnaire (MSQ) \cite{Gianaros2001-ny}. The MSQ is an Ethics Board requirement and part of the standard simulator operating procedure to ensure participants’ wellbeing before continuing the research drive, and does not form part of the study design for further analysis.

\textbf{Research Drive:} After a short break the research drive started and took ca. 15 minutes. In this, automated driving was activated from the beginning. The participants needed to respond to markers throughout the drive and continue watching the video clips over the NDRA display, while remaining ready to continue driving at any point. 

\textbf{Questionnaires, Interview and Debriefing:} After finishing the drive, participants filled out the questionnaires, had a semi-structured interview, received their vouchers for their participation, and were debriefed. Participation took approximately 1.5 hours overall. 

\section{Results}
In the following, we discuss the findings of the heatmap observations in relation to the graph of the NDRA engagement between the treatments, the system usability score (SUS) and the NDRA recall performance.

\subsection{Gaze Behaviour}
To explore differences in eye gaze behaviour between the four groups, we compared them by 1) plotting the percentage of participants within each group fixated in the NDRA smoothed over time, 2) creating a heatmap video visualisation of fixation densities for each frame, and 3) a thematic analysis of the resulting heatmap video visualisations. The gaze data of 6 participants had technical issues during the recording, leading to calibration and synchronisation issues, and were therefore excluded from the gaze behaviour analyses. The gaze behaviour results presented are based on the remaining N=209.

\subsubsection{NDRA Fixations Over Time}
\begin{figure*}[ht]
  \centering
  \includegraphics[width=\textwidth]{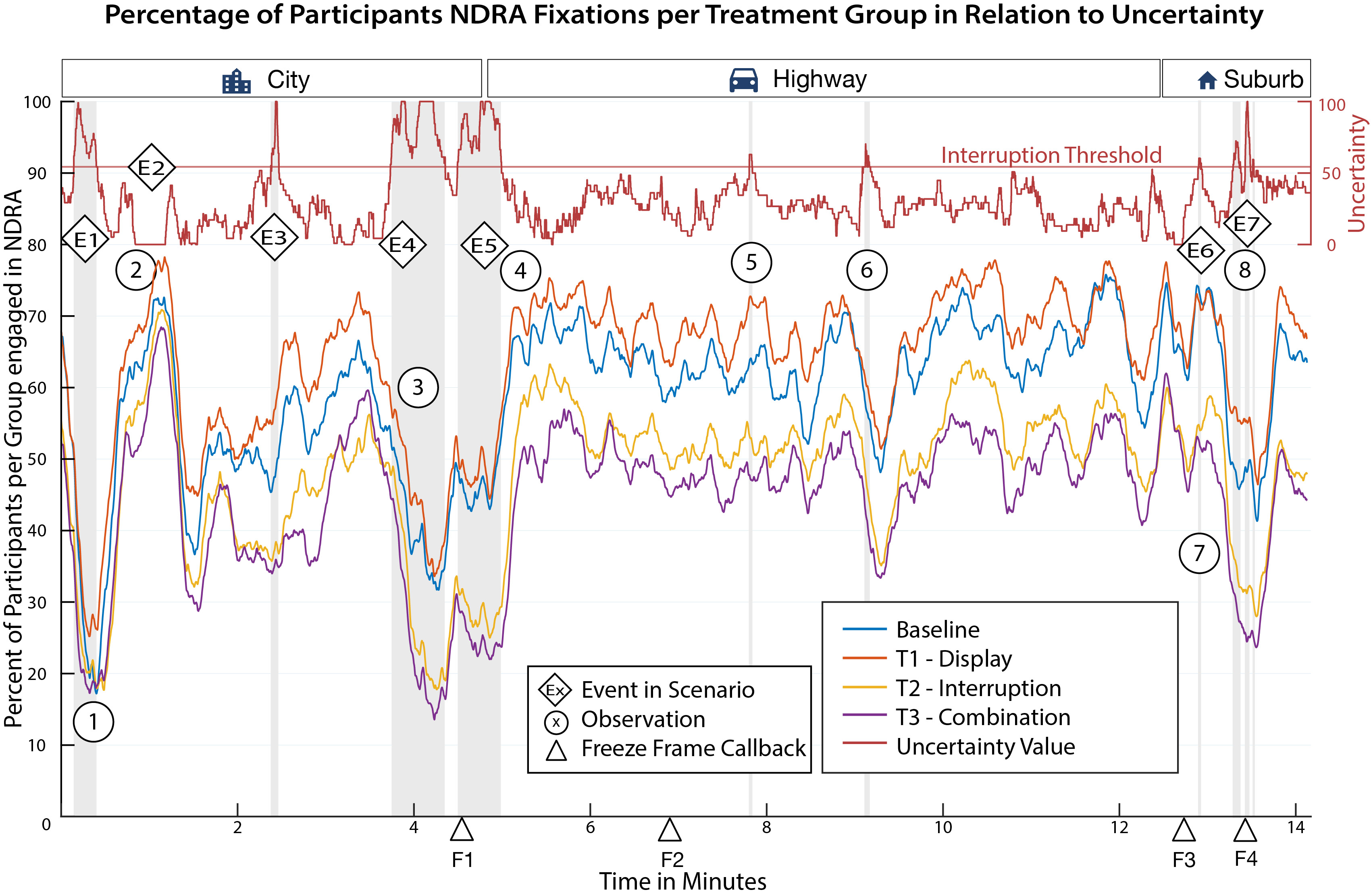}
 \caption{ The graph shows the driving scenario with its distinctive events (diamonds 1-7), quantified uncertainty (top red graph, see Quantifying Uncertainty section), and the freeze frame moments (triangle F1-F4; for the post-trial assessments, see Freeze Frame section) over time. The percentage of participants per group engaged in the NDRA is shown relative to the uncertainty and interruptions (grey areas).}
 \Description{A graph shows the uncertainty of the study drive. It's divided into three sections, the city drive, the highway and suburb. On the x-axis is the time, and on the y-axis is the uncertainty. }
 \label{fig:NDRAEngagementGraph}
\end{figure*}

 Figure \ref{fig:NDRAEngagementGraph} shows the percentage of participants within each condition that fixated on the NDRA percentage (y-axis) over time (x-axis). It is visualised against the quantified uncertainty (cf. Quantifying Uncertainty section above) on top, including the threshold for the NDRA interruption, to provide the context of when NDRA interruptions occurred and how the situations escalated towards interruptions. We calculated the percentage for each frame, and then smoothed it over time in Matlab with the function \textit{movemean} using a time window of 450 frames (15 seconds). This time frame was used to visually inspect patterns in the data and identify broader patterns over longer time intervals while balancing noise and signal. The time window of 15 seconds was identified through testing as the shortest time window to still observe these patterns throughout the 14-minute-long drive before becoming too noisy or messy. The times of interruptions are visualised with a grey background.

\subsubsection{Gazes Towards the NDRA}
To investigate the effects of the treatments on the overall duration of gazes in the NDRA display (Table \ref{tab:DescriptiveStatsNDRAGazesEntireDrive}), a one-way between-groups ANOVA was conducted using the four conditions. There were no extreme outliers, as assessed by boxplot; data was normally distributed for each group, as assessed by Shapiro-Wilk test (p > .05); and there was homogeneity of variances, as assessed by Levene's test of homogeneity of variances (p = 0.153). NDRA gaze duration was found to be significantly different between treatment groups \( F(3, 205) = 10.29, p < .001 \). Post-hoc pairwise comparisons using Tukey's HSD test were performed and revealed significant differences between several treatment pairs. These are described in Table \ref{tab:CompairsonNDRAGazesEntireDrive}. 

\begin{figure}[ht]
 \centering
 \includegraphics[width=\linewidth]{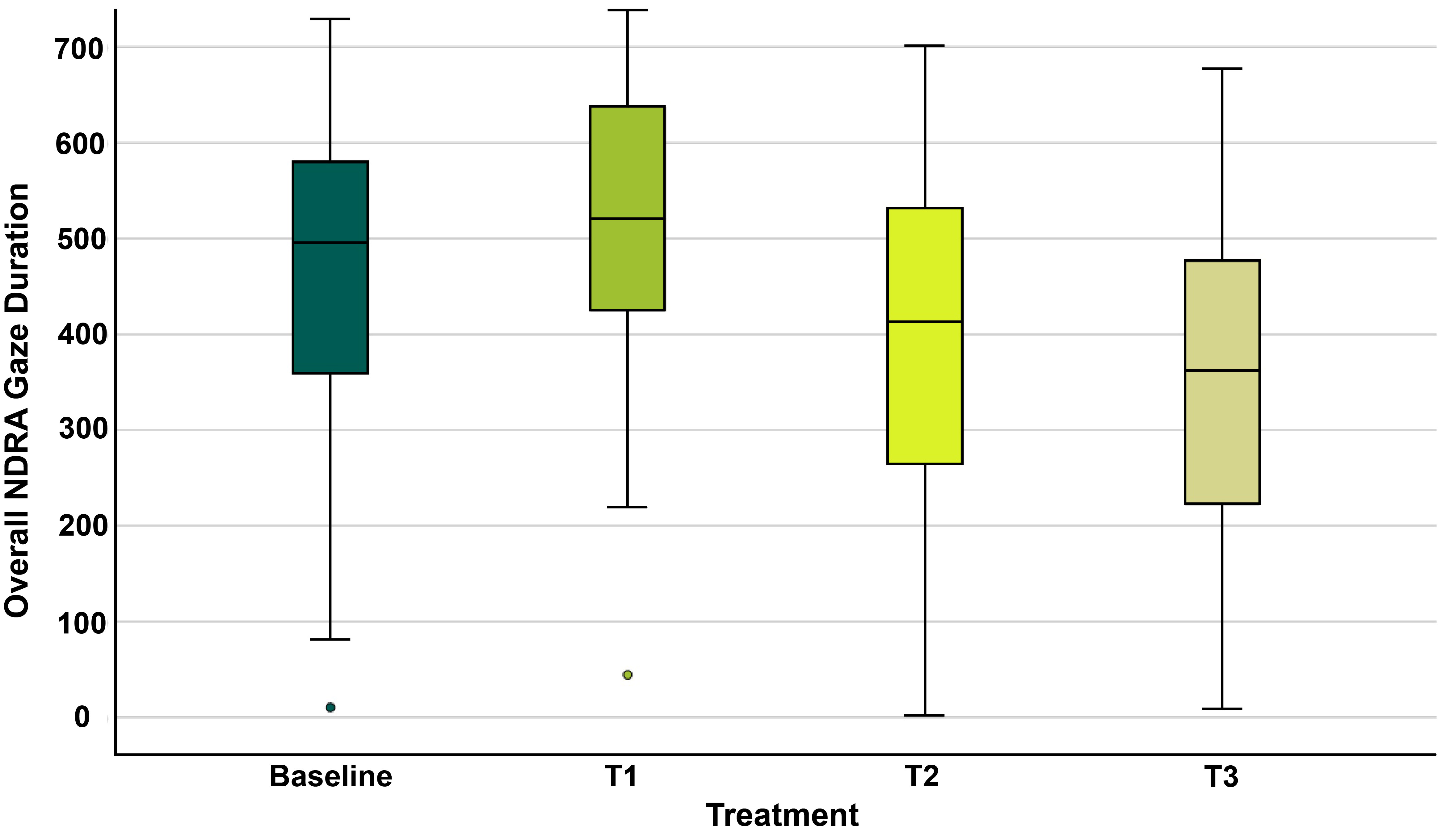}
 \caption{Boxplot of the overall gaze duration between participants in seconds}
 \Description{Boxplot of the descriptive results above}
 \label{fig:BoxPlotGazeOverall}
\end{figure}

The boxplot visualisation of NDRA engagement across treatments (Figure \ref{fig:BoxPlotGazeOverall}) illustrates these findings, highlighting the median, interquartile ranges, and moderate (less than 3 box lengths) outliers within each treatment group.

\begin{table}[ht]
\centering
\caption{Descriptive statistics of overall gaze duration in the NDRA display in seconds}
\label{tab:DescriptiveStatsNDRAGazesEntireDrive}
{\footnotesize
\begin{tabular}{cccccccccc}
\toprule
Group & N & Mean & SD & Min & 25\% & 50\% & 75\% & Max \\
\midrule
BL & 52 & 473.25 & 155.32 & 10.27 & 360.77 & 495.60 & 577.73 & 729.20 \\
T1 & 50 & 511.81 & 152.74 & 44.40 & 425.83 & 520.77 & 635.78 & 750.73 \\
T2 & 54 & 376.47 & 195.27 & 1.67 & 265.45 & 413.27 & 530.31 & 701.10 \\
T3 & 53 & 354.53 & 169.55 & 9.07 & 222.83 & 362.43 & 476.40 & 677.27 \\
\bottomrule
\end{tabular}
}
\end{table}

\begin{table}[ht]
\centering
\caption{Tukey’s HSD for gazes in NDRA}
\label{tab:CompairsonNDRAGazesEntireDrive}
\begin{tabular}{ccccccc}
\toprule
G. 1 & G. 2 & Mean Diff & p-adj & Lower & Upper & Reject \\
\midrule
BL & T1 & 38.56 & 0.6413 & -48.40 & 125.52 & False \\
BL & T2 & -96.77 & 0.0191 & -182.08 & -11.47 & True \\
BL & T3 & -118.71 & 0.0023 & -204.41 & -33.02 & True \\
T1 & T2 & -135.33 & 0.0010 & -221.50 & -49.17 & True \\
T1 & T3 & -157.27 & 0.0010 & -243.83 & -70.72 & True \\
T2 & T3 & -21.94 & 0.9000 & -106.83 & 62.95 & False \\
\bottomrule
\end{tabular}
\end{table}

\subsubsection{Gazes in the NDRA Display during Low Uncertainty (not interrupted)}
\begin{figure}[ht]
 \centering
 \includegraphics[width=\linewidth]{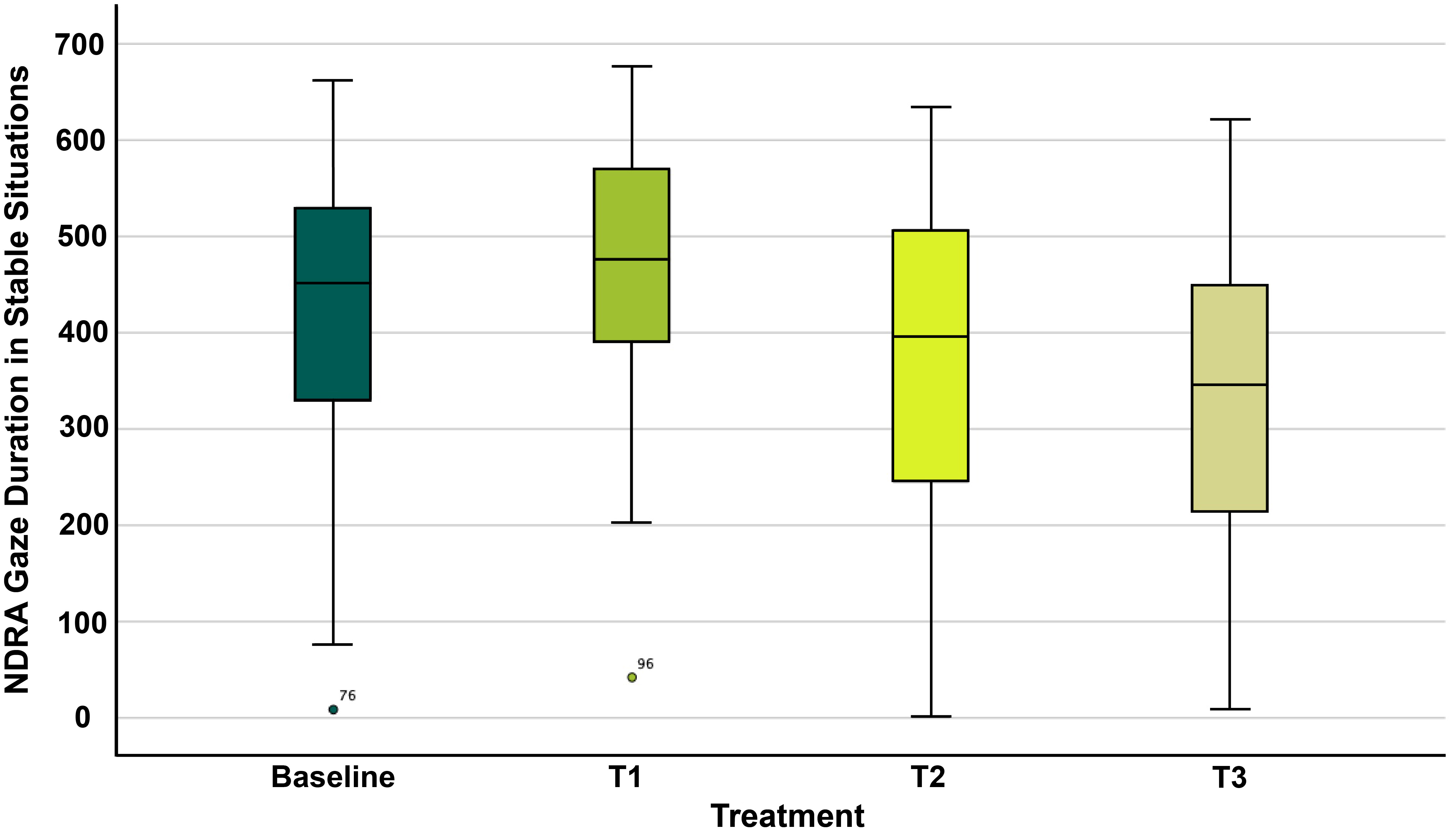}
 \caption{Boxplot of the overall gaze duration between participants in seconds}
 \Description{Boxplot of the descriptive results above}
 \label{fig:BoxPlotGazeStable}
\end{figure}

Here we investigate the effects during low-medium uncertainty driving situations—i.e. excluding times during which participants were interrupted in T2 and T3 and the uncertainty Display was high/red in T1 and T3— and only including times during which the video was playing in the NDRA Display in all conditions (Table \ref{tab:DescriptiveStatsNDRAGazesStable}).
A one-way between-groups ANOVA was performed using the four conditions (with sub-samples as described). There were no extreme outliers, as assessed by boxplot; data was normally distributed for each group, as assessed by Shapiro-Wilk test (p > .05); and there was homogeneity of variances, as assessed by Levene's test of homogeneity of variances (p = 0.099). NDRA gaze duration was found to vary between groups in low-medium uncertainty driving situations \(F(3,205)=8.71,p<.001\). Post-hoc pairwise comparisons using Tukey's HSD test were performed and revealed significant differences between several treatment pairs. These are described in Table \ref{tab:CompairsonNDRAGazesStable}, consistent with the previous analysis, significant differences were found for NDRA gaze duration such that NRDA gaze duration was lower in T2 than baseline and T2, lower in T3 than baseline, lower in T2 than T1, and lower in T3 than T1). The boxplot visualisation (Figure \ref{fig:BoxPlotGazeStable}) illustrates these findings, highlighting the median, interquartile ranges, and moderate (less than 3 box lengths) potential outliers within each treatment group. Compared to the baseline, the boxplot visualises the tendency \footnote{statistically not significant} for the uncertainty display (T1) to increase NDRA engagement, while both interruption groups (T2 and T3) show significantly reduced NDRA engagement, even if they are not interrupted (see Figure \ref{fig:BoxPlotGazeOverall} and \ref{fig:BoxPlotGazeStable}). While the uncertainty display increased the NDRA engagement, the combination with managed interruptions had the opposite tendency to decrease the NDRA engagement.

\begin{table}[ht]
\centering
\caption{Descriptive statistics of overall gaze duration in the NDRA Display in low-medium uncertainty situations}
\label{tab:DescriptiveStatsNDRAGazesStable} 
{\footnotesize
\begin{tabular}{cccccccccc}
\toprule
Group & N & Mean & Std & Min & 25\% & 50\% & 75\% & Max \\
\midrule
BL & 52 & 429.60 & 139.38 & 8.53 & 331.45 & 451.37 & 526.88 & 662.23 \\
T1 & 50 & 465.40 & 136.90 & 42.03 & 391.61 & 476.40 & 569.18 & 676.70 \\
T2 & 54 & 351.65 & 179.88 & 1.07 & 245.25 & 396.15 & 502.12 & 634.93 \\
T3 & 53 & 331.71 & 155.85 & 8.57 & 213.97 & 345.90 & 449.50 & 621.70 \\
\bottomrule
\end{tabular}
}
\end{table}

\begin{table}[ht]
\centering
\caption{Tukey’s HSD for gazes overall in NDRA in low-medium uncertainty situations}
\label{tab:CompairsonNDRAGazesStable} 
\begin{tabular}{ccccccc}
\toprule
G. 1 & G. 2 & Mean Diff & p-adj & Lower & Upper & Reject \\
\midrule
BL & T1 & 35.80 & 0.6289 & -43.42 & 115.02 & False \\
BL & T2 & -77.95 & 0.0490 & -155.65 & -0.24 & True \\
BL & T3 & -97.88 & 0.0074 & -175.95 & -19.82 & True \\
T1 & T2 & -113.75 & 0.0013 & -192.24 & -35.25 & True \\
T1 & T3 & -133.69 & 0.0010 & -212.54 & -54.83 & True \\
T2 & T3 & -19.94 & 0.9000 & -97.27 & 57.40 & False \\
\bottomrule
\end{tabular}
\end{table}

\subsubsection{Eye Gaze Heatmap Video}
A heatmap video was created from the eye gaze data in Matlab with a customised version of \textit{Auto Heat Mapper}\footnote{\url{https://github.com/aulloa/Auto_Heat_Mapper}, retrieved September 2023} that uses the Bivariate Kernel Density Estimation (gkde2). The calculated gkde-maps were then overlaid with the original video frames. The frames from the four groups were synced and collated into one video. This video\footnote{Available in its entirety as supplementary material of this paper} shows time-synced moving eye gaze heatmaps - overlaid over the driving and intervention context - from all four groups simultaneously for direct comparison (Figure \ref{fig:Result1}). 

\subsubsection{Observations \& Thematic Analysis of Quantified Eye-Gaze Heatmap Video}
The thematic analysis was done by the first author in an iterative manner in consultation with the research team. He watched the video repeatedly from start to end in slow motion to identify visible gaze behaviour differences between each group in the heatmaps video. Those differences were noted down with accompanying screenshots, organised in a timeline, discussed and then similar observations clustered into themes. The themes provide a more nuanced understanding of the gaze pattern differences observed in statistical analyses above: \newline

\textbf{Similar Starting Behaviour:}\newline
At the beginning of the data collection drive, the four groups showed similar eye-gaze heatmap distributions over time, i.e. collectively they appeared to gaze and switch between watching the entertaining movie clips on the NDRA screen and monitoring the driving environment in similar ways. This is also reflected at the beginning of the graph (Figure \ref{fig:NDRAEngagementGraph}). Their responses to triggers in the driving environment, such as other road users, were observed similarly between all four groups. This trend changed after 01:20 min (2) with a stop at the red traffic light.\newline

\begin{figure*}[ht]
  \centering
  \includegraphics[width=\textwidth]{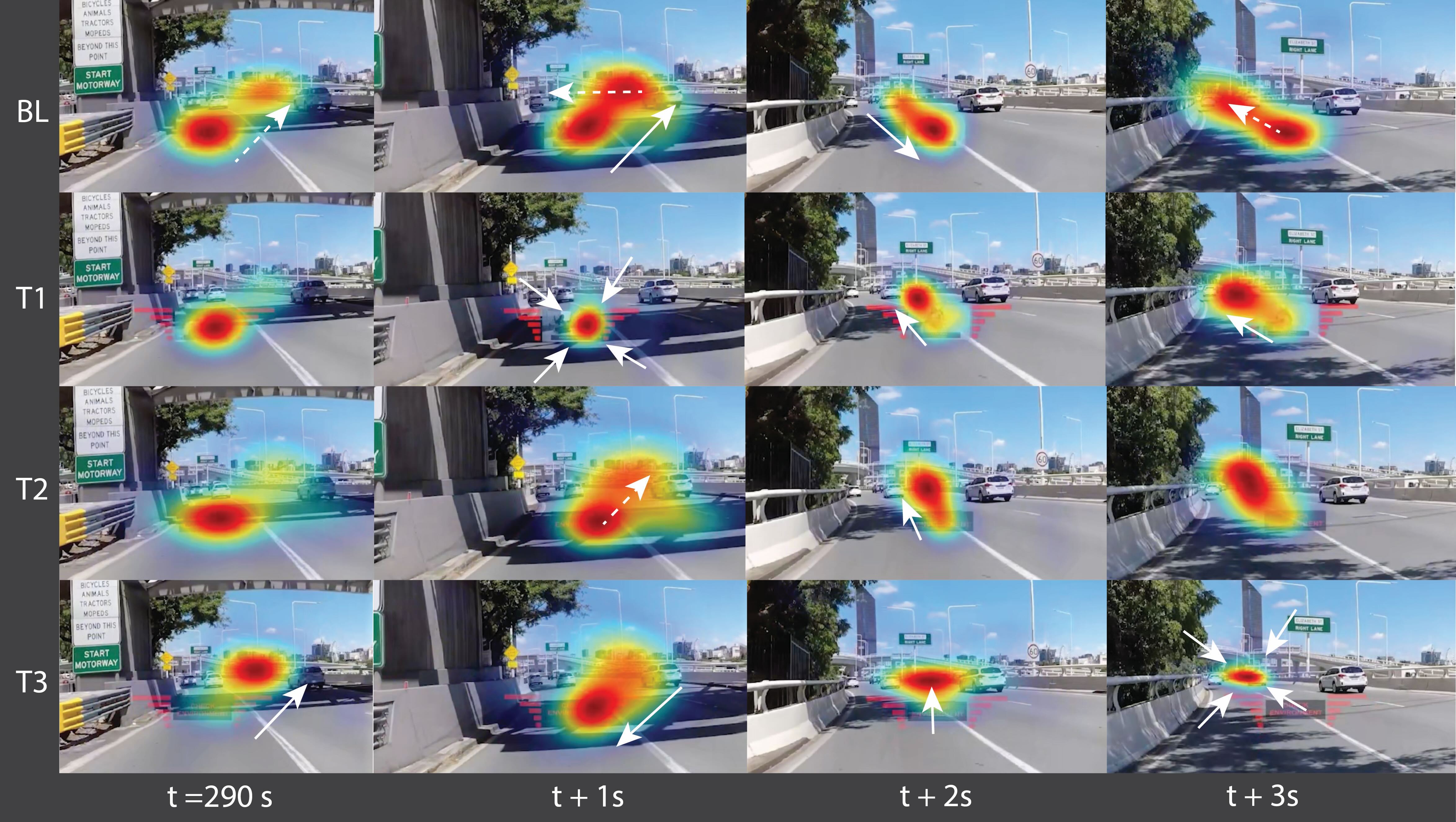}
 \caption{ 
 Screenshots of eye-gaze heatmap evolving over time (x-axis) during highway merging (high uncertainty), highlighting differences in gaze behaviour between groups (y-axis). Full arrows indicate the completed movement of "heat" and dotted arrows indicate emerging movement. T3 moving earliest away from NDRA Display (t=290s), T1 remaining focused on NDRA Display (t+1s)}
 \Description{This figure shows 4 times the same driving scene in a grid. Each of the driving scenes includes a different display intervention that supports fallback readiness, described in the intervention section. Overlayed on top of it are the eye-gazes of the participants in form of a heatmap. The details are described in the discussion}
 \label{fig:Result1}
\end{figure*}

\textbf{Greater Gaze Dispersion in T2-Interruption and T3- Combination:}\newline
We observed that participants in the T1-Display group had a more focused, smaller heat area, whereas both T2-Interruption and T3-Combination had a wider spread of eye gazes throughout the drive. Two examples show a situation while entering (Figure \ref{fig:Result1}) or exiting the highway, with the developing hazard (Figure \ref{fig:teaser}). Unexpectedly, we observed this “focus” effect more often when uncertainty was high.\newline

\textbf{NDRA Focus in T1-Display:}\newline
We observed that the eye-gaze fixations of the T1-Display group would typically remain longer in the area of the NDRA display compared to the other groups (BL, T2-Interruption, T3-Combination). E.g., in T1 compared to the other groups, we could observe a delayed shift of eye-gazes towards areas of interest, such as other road users, street signs or gazing in the future pathway of a curved road. This observed discrepancy worsened over time and culminated in an obvious late response to the more safety-critical event of the emerging hazard at the end of the drive. \newline

\begin{figure*}[ht]
  \centering
  \includegraphics[width=\textwidth]{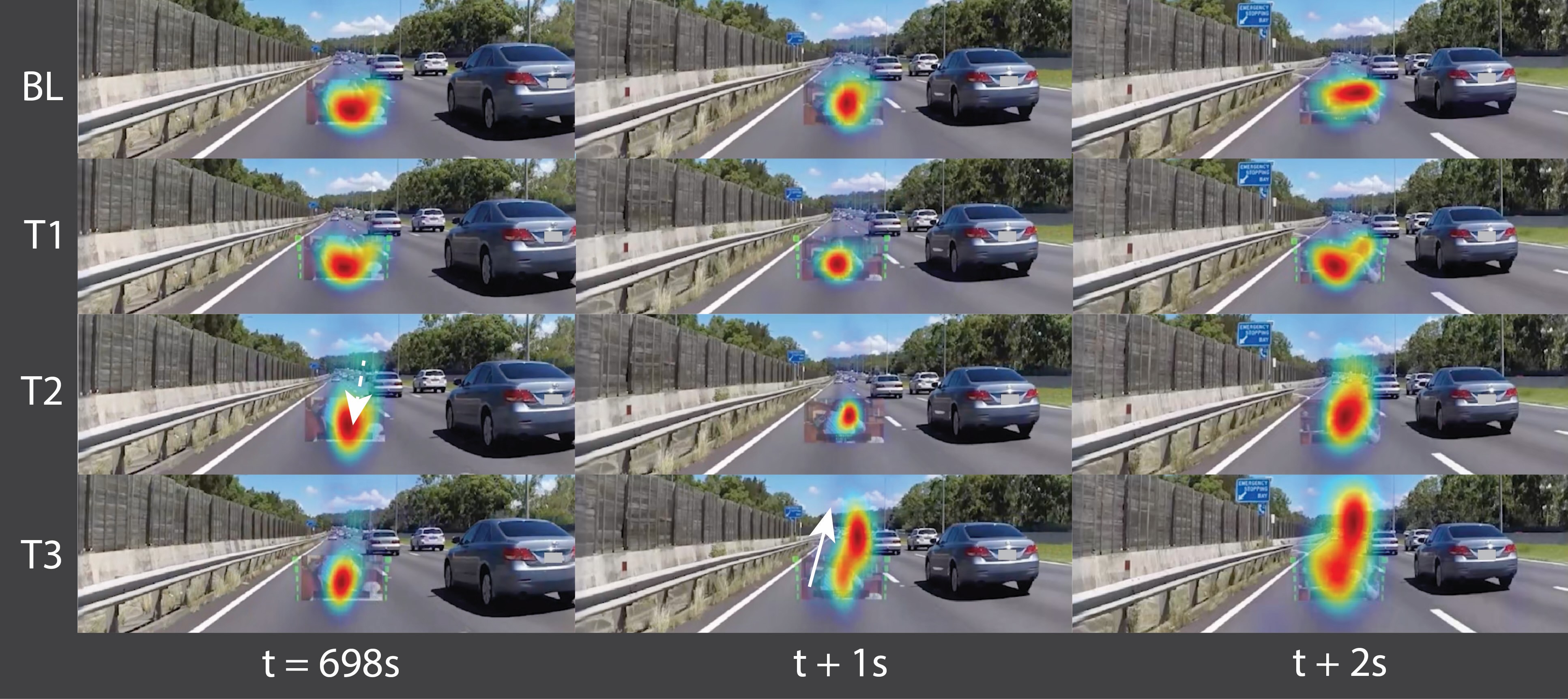}
 \caption{ Screenshots of eye-gaze heatmap evolving over time (x-axis) during highway driving (low uncertainty), highlighting differences in gaze behaviour between groups (y-axis). Full arrows indicate the completed movement of "heat" and dotted arrows indicate emerging movement. T2 and T3 show monitoring effects (eye-gaze to a vehicle ahead) that appear amplified in the combination group.}
 \Description{This figure shows 4 times the same driving scene in a grid. Each of the driving scenes includes a different display intervention that support fallback readiness, described in the intervention section. Overlayed on top of it are the eye-gazes of the participants in form of a heatmap. The details are described in the discussion}
 \label{fig:Result2}
\end{figure*}

\textbf{Monitoring Echos in T2-Interruption and T3-Combination:}\newline
In both interruption groups (T2 + T3), we observed an effect that we describe as “echoes” of the interruption treatment. Following an interruption and the NDRA being displayed again, both groups had repeated, rhythmic and synchronous gaze towards the driving environment (Figure \ref{fig:Result2}). Those echoes are best observable after entering the highway and continuing throughout the highway drive, during which overall fewer interruptions occurred due to the nature of highway driving being a more stable driving environment (Figure \ref{fig:NDRAEngagementGraph}, between 4 and 7) compared to driving on urban or suburban roads. The echo effect appeared to be amplified in frequency and strongest (in terms of “heat”) in the Combined Group (T3), and could not be observed in either of the non-interruption groups (BL, T1). \newline

\textbf{Highest visual attention towards driving task in T3:}\newline
We observed both groups with interruptions (T2 + T3) fixated more often on events in the driving environment (Figure \ref{fig:Result2} and \ref{fig:teaser})—such as overtaking or undertaking vehicles, passing street signs — by directing their gaze towards them. Comparing just the two interruption groups, the responses in the Combined Group (T3) were observable faster and appeared more often. Unexpectedly, these responses occurred even outside of the rhythmic echoes described in the previous paragraph, and in between high uncertainty situations. The graph in Figure \ref{fig:NDRAEngagementGraph} corroborates this observation in that it shows that T3 had the lowest percentage of participants engaged with the NDRA.\newline

\textbf{Response to Uncertain Situations:} \newline
Both interruption groups respond differently in uncertain situations. Especially towards the end of the drive where a few shorter intermittent bursts of high uncertainty (cf. Figure \ref{fig:NDRAEngagementGraph}) in close succession associated with the vehicle preparing to exit the highway, changing lanes to exit and the emerging hazard shortly after merging on the main road. Both interruption groups were correspondingly briefly interrupted in close succession. Observations from this phase suggest that participants in T2-Interruption were more likely to briefly direct their gaze back to the NDRA Display when the video returned, whereas participants in T3-Combination were more likely to keep their gaze locked on the hazard (Figure \ref{fig:teaser}). Lastly, it was noticeable when comparing T2 and T3 during this phase, that participants in T2-Interruption were more likely to stay slightly longer in the NDRA Display before shifting their gaze towards monitoring the driving scene again when the next interruption was triggered. 

\subsection{Complementary Measures}
\subsubsection{System Usability Scale (SUS)}
Participants subjectively evaluated the overall automated driving system usability of the automated driving system after the data collection drive. One participant survey was lost during the data collection because of a technical error. Several outliers were identified through the assessment of boxplots. These outliers were considered valid data, so the decision was made to assess the data using non-parametric statistics rather than excluding cases or transforming the data. A \textbf{Kruskal-Wallis H-test} was conducted to determine if there were differences in the System Usability Scale (SUS) scores between groups \textbf{Baseline (BL)} (N=53), \textbf{T1-Display} (N=54), \textbf{T2-Interruption} (N=53) and \textbf{T3-Combination} (N=54). The distribution of the overall SUS scores was similar for all groups assessed by visual inspection of a boxplot. The Medians are in the \textbf{Baseline} (55.71), \textbf{T1-Display} (56.43), \textbf{T2-Interruption} (55.71) and \textbf{T3-Combination} (56.43), but the differences were \textbf{not statistically significant} X\textsuperscript{2}(3)=1.543, p=.672.

\subsubsection{NDRA Recall Performance}
The analysis of the NDRA Recall Performance focuses on the participants' wrong answers in the post-trial questionnaire. Several outliers were identified through the assessment of boxplots. Outliers are defined as deviations 1.5 times larger or lower than the interquartile range (for all reported results). These outliers were considered valid data, so the decision was made to assess the data using non-parametric statistics rather than excluding cases or transforming the data. A \textbf{Kruskal-Wallis H-test} was conducted to determine if there were differences in the NDRA Task Engagement survey between the treatments \textbf{Baseline (BL)} (N=51), \textbf{T1-Display} (N=53), \textbf{T2-Interruption} (N=53) and \textbf{T3-Combination} (N=54). The distribution of the results was similar for all groups assessed by visual inspection of the boxplot. The median of wrong answers resulted in the \textbf{Baseline} (1), \textbf{T1-Display} (2), \textbf{T2-Interruption} (2) and in \textbf{T3-Combination} (2). However, the differences were \textbf{not statistically significant} X\textsuperscript{2}(3)=5.419, p=.144.

\subsubsection{Marker Click Response}
To assess the impact of the treatments on Marker Click Response, we looked at two outcome variables; the success rate (i.e., the percentage of time participants noticed the marker and clicked) and reaction time (i.e., how quickly they indicated noticing the marker). For success rate, the assumption of normality was violated as evidenced by the Shapiro-Wilk test (p < 0.05 for all groups). Hence, rather than employing ANOVA, we analysed the difference between treatments using a Krusal-Wallis test. No significant difference was found for success rate across conditions (H = 5.246, p = 0.155). For reaction time, no violations of normality were found, however the assumption of homogeneity of variance was not met (Levene’s test, =0.462). Hence, we analysed the difference between treatments using a Kruskal-Wallis test. No significant difference was found for response time across conditions (H = 5.246, p - 0.155).

\section{Discussion}
Our study investigates how HMI interventions based on communicating uncertainty influence the eye-gaze behaviour of fallback-ready users when engaged in everyday non-driving related activities (NDRAs). This discussion is split into three themes based on results from statistical and heatmap analyses. 

\subsection{Designing NDRA interruptions may benefit fallback-readiness}
We have observed \textbf{continuous rhythmic—or “echoing” — monitoring gazes in both interruption groups} (T2+T3) that started appearing after entering the highway and continued throughout the drive. I.e. both interruption groups (T2+T3) regularly directed their gaze away from the NDRA to check the environment between NDRA interruptions.  Those findings are confirmed by the inferential statistics reports in 5.1.2 and 5.1.3.

These “echos” could be explained in that the message to “check environment”, which interrupted the NDRA, served as a timely reminder to the participants of their responsibilities to remain fallback-ready and to monitor the system and environment. However, whether users engage in such cognitive processes would need to be explored further.

Another explanation could be grounded in the task interleaving stage as described by Janssen et al. \cite{Janssen2019-ib}, who explore this concept in the context of a full control transition, where fallback-ready users task-switch from NDRA to the full control of the vehicle. During this task interleaving stage users intuitively and temporarily interleave their attention between the NDRA and the (driving) monitoring task. Merat et al. \cite{Merat2014-wg} found that more predictable and system-based disengagements (and therefore full transitions) increased subsequent intermittent attention towards the road centre when the automation was active again. 

Our study shows that this kind of task interleaving can be observed (and designed for) outside of a full control transition. In our interruption groups (T2+T3), participants were nudged towards a partial transition to the role of a “passive observer”, for which a smaller task-set reconfiguration is required, in particular, if the reconfiguration is guided or supported by the system as done in the interruption groups. These externally triggered partial transitions then appear to subsequently lead users to a temporary interleaving of NDRA and monitoring, which we describe as monitoring echoes, and which Janssen et al. refer to as voluntary “self-interruptions” devoid of explicit external triggers \cite{Janssen2019-ib}.

The timing and persistence of those echos—i.e. when those self-interruptions occur and when they start to occur less regularly—will need to be explored further. The observed synchronous timing of the monitoring gazes in T2 and T3 hints at aspects of Wicken’s SEEV Model of Visual Attention Allocation \cite{Wickens2001-pm, Wickens2003-mn}, and could be linked to the saliency or value within the NDRA, e.g. natural breakpoints such as the end of a sentence or change of scene, or saliency or expectancy in the driving environment.

While the findings of having significantly more glances outside of the NDRA in T2 and T3 even when not interrupted (c.f. 5.1.3) seem promising in terms of increasing situation awareness, the effect and implications these echos have on (i) improving users’ fallback-readiness (in particular in terms of takeover performance and safety measures), (ii) visual and cognitive effort to stay in the loop, (iii) the acceptability of the overall AV system, (iv) compliance with roles and responsibilities, and (v) complacency and bias towards the automated system \cite{Endsley2017-qd, Kunze2019-gj, Llaneras2017-sw,Parasuraman2010-mw}, all need to be investigated further. Our study showed no significant difference in detecting marker clicks; however, the markers were primarily added to facilitate a dual-task competition rather than a proxy measure for situation awareness.

\subsection{Beware of the utility, meaning and effect of peripheral system information displays}
We discovered multiple indicators that adding the uncertainty display design—the “Guardian Angel”, which represents a peripheral display of system information—may have a different utility for users depending on the overall HMI context or concept it is utilised in. This difference in utility can be derived from comparing the effects of the addition on fallback-ready users’ gaze behaviour. 

It further appears that in T1-Display, the meaning of the peripheral display represents merely a soft and often ignored \textbf{“suggestion”} to check the environment. In contrast, in T3-Combination, where the peripheral display is used in combination with interruptions, the display serves as a kind of \textbf{“pre-warning”} to a subsequent enforced command to check the environment.

The mere \textbf{“suggestion”} to check the environment (T1) may not be sufficient if participants feel burdened by having to decide for themselves whether to interrupt their NDRA, all the while being engaged in and distracted by it. This could explain why fewer participants monitored the driving environment. Resolving the internal conflict of the dual task between attending to the NDRA and being suggested to interrupt it, could lead to misjudgment of the current uncertainty situation, especially on top of an increasingly unstable situation. This poor decision-making (or conflict resolution) performance could be explained by the bad risk estimation abilities of humans in dual-task conditions \cite{Horrey2009-aj, Horrey2009-xf, Janssen2012-cq}. It should be noted that the performance is influenced by motivational conditions \cite{Kim2015-eg, Marberger2018-cm} to interrupt the desired task, such as the NDRA. In essence, the negotiation process—so “voluntarily” interrupting a desired entertaining task due to a) external stimuli and/or b) being triggered by the responsibility of remaining fallback-ready to comply with the monitoring obligations—might add additional cognitive workload to solve a driving problem. Such considerations, however, were outside the scope of this paper and should be investigated in future research.

From our observations, we argue a positive effect of \textbf{“pre- warning”} an interruption through an uncertainty display. In both interruption groups (T2, T3), the user is freed of the burden of deciding whether to interrupt the NDRA and have an improved performance, but between the two, T3 appears to interrupt the NDRA faster and more often than T2. We assume the uncertainty display makes the interruption more transparent and predictable, which could be explained by the principle of graceful degradation \cite{Hancock2019-qf, Janssen2019-ib}. Announcing a potential future task switch allows the user to mentally prepare for it as the system as a whole becomes more predictable. This has been shown to lead to reduced workload, and increased (monitoring) task performance and usability \cite{Monsell2003-an, Monsell2003-oy, Van_der_Heiden2017-gt}.

Our study hints that the utility, the meaning and the effect of uncertainty displays need to be carefully considered (and tested) in the context of the overall HMI concept. However, bringing this finding into the bigger picture, we recommend designers of HMIs test their interventions also in context because reciprocal relations can change the context of an intervention.

\subsection{Do not be afraid to interrupt the NDRA}
We did not find a significant difference in a) the NDRA performance by asking questions about the content of the videos, nor b) subjective ratings of the System Usability Scale (SUS). A reduced NDRA performance could indicate a higher SA of the driving activity \cite{Schomig2013-ds, Kunze2019-gj, Beller2013-ir}. Our results indicate that the interventions allow following less performance-critical NDRAs like watching movies. However, the NDRA performance was not the focus of our study. Participants were primed to pay attention to the videos as part of creating the dual-task competition so they would not simply ignore it. Our results indicate that participants across all four conditions may have paid similar attention to the NDRA. Regarding usability, our results indicate that the interventions did not negatively impact it, even though one could have expected that interrupting the NDRA would have made a difference. 
 
To summarise the results, the different interventions did not significantly impact what participants remembered about the video they watched as the NDRA or the usability. This shows that fallback-ready users might accept heavy-handed design interventions in their entertaining NDRAs even if it is interrupted from time to time. In-vehicle HMI designers should not be afraid to do so in the context of automated driving. 

\section{Limitations}
Even though the study was conducted in a high-fidelity simulator with a motion platform, there was no simulated motion (celeration) during the drives, except for a subwoofer under the driver's seat to simulate vehicle vibrations. Research shows that motion influences the drivers' or fallback-ready users' behaviour in automated driving \cite{Sadeghian_Borojeni2018-kh}. The focus and scope of this study was to align the eye-gazes accurately with the driving environment on the video screens that are not connected to the motion platform. A vehicle in motion would have complicated alignment beyond the scope of this study. However, the condition of not having motion was consistent throughout the drive, and all participants, independent of their treatment, experienced the driving scenario the same way. Individual differences between subjects may be an alternative explanation for results, although this was mitigated through a large sample size and balanced distribution (see Table 1).

The HUD was simulated as a transparent video overlay on the simulation screen, i.e. at the same depth. As such, our setup can not differentiate between participants watching the content of the NDRA, or checking the driving environment “behind” the HUD. We can only assume that eye gazes directed towards the HUD suggest watching the NDRA rather than checking the driving environment. This assumption is grounded in Wicken’s SEEV Model of Visual Attention Allocation \cite{Wickens2001-pm, Wickens2003-mn}.

Research has shown that more engaging and challenging tasks have higher impacts on fallback readiness \cite{Jarosch2019-mi}. In our study, the NDRA was to watch entertaining videos paired with a memory task to motivate paying attention. As such, our findings may not apply to NDRAs that are mentally more/less demanding or more/less visually distracting.

Lastly, the click response task to react to markers in the environment could be considered a third competing task, in addition to the dual task competition (monitoring and remaining ready to continue driving versus engaging in an entertaining NDRA) that is being studied here. This response task may also pose an additional distraction, particularly in the situation of an emerging hazard. We argue the lack of risk from the simulation environment justifies this incentivisation of monitoring the driving task and note that all participants had the same markers and incentivisation throughout the study. 

\section{Conclusion}
This publication investigates changes in eye gaze behaviour through uncertainty design interventions in conditional automated vehicles (SAE Level 3) and above, while fallback-ready users are engaged in everyday non-driving related activities (NDRAs). We have observed differences between the treatments of a visual uncertainty display (T1, T3) and an interruption of the NDRA (T2, T3). More precisely, we discovered that the function of visual uncertainty displays changes with interruptions. We suggest researchers test design interventions in the context of other interventions to explore if the interventions interact with others and might change their intended purpose.

In the context of this research, we discovered that uncertainty visualisation in combination with everyday NDRA might reduce the monitoring fixations and might lead to misjudgment of the situation. We further found that interruptions based on uncertainty might lead to task-interleaving between NDRA and monitoring. Combining the interruptions with an uncertainty visualisation seems promising in creating a broader task-interleaving -behaviour with more frequent and voluntary interruptions of the NDRA. This increased number of fixations in the driving environment from the task interleaving could lead to a higher SA and therefore improved fallback readiness and safety performance, which needs to be investigated further in the future.

\begin{acks}
  This work is funded from an Australian Research Council’s Linkage Project funding scheme (project number LP150100979). We thank Seeing Machines Ltd. for their support of the project and for making the Driver Monitoring System (DMS) available in our Advanced Driving Simulator. Also, a special thanks to SensoDrive GmbH, Dr. Xiaomeng Li, Alice Conant, Mohammad Faramarzian, Wanda Griffin, Adrian Wilson, Dr. Sebastien Demmel, Dr. Sepehr Ghasemi Dehkordi, Jorge Luis Prado Gaytan, Julia Vehns and Kim Smith.
\end{acks}
\bibliographystyle{ACM-Reference-Format}
\bibliography{chi24329.bib}
\end{document}